\documentclass[twocolumn]{aastex63}

\usepackage[utf8]{inputenc}
\usepackage{amssymb,amsmath,graphicx,latexsym}
\usepackage{array}
\usepackage{bm}
\usepackage{color}
\usepackage{dcolumn}
\usepackage{multirow}
\usepackage{float}
\usepackage[utf8]{inputenc}
\usepackage{url}

\usepackage{ulem}
\usepackage{grffile}
\usepackage{extarrows}
\usepackage{tikz}
\usepackage[caption=false]{subfig} 
\usepackage{hyperref}

\newcommand{\HI}{\textsc{Hi}\,}
\newcommand{\bx}{\boldsymbol{x}}
\newcommand{\bs}{\boldsymbol{s}}
\newcommand{\vf}{\boldsymbol{f}}
\newcommand{\vt}{\boldsymbol{t}}

\newcommand{\bR}{\boldsymbol{R}}

\newcommand{\bS}{\boldsymbol{S}}
\newcommand{\bW}{\boldsymbol{W}}
\newcommand{\bI}{\boldsymbol{I}}

\newcommand{\bU}{\boldsymbol{U}}
\newcommand{\bD}{\boldsymbol{D}}
\newcommand{\bDelta}{\boldsymbol{\Delta}}
\newcommand{\bxi}{\boldsymbol{\xi}}
\newcommand{\bQ}{\boldsymbol{Q}}
\newcommand{\bLambda}{\boldsymbol{\Lambda}}

\DeclareMathOperator{\Tr}{Tr}

\shorttitle{eGNILC foreground removal}

\shortauthors{Dai \& Ma}

\begin{document}

\title{Expanded Generalized Needlet Internal Linear Combination (eGNILC) Framework for the 21-cm Foreground Removal}


\author{Wei-Ming Dai}

\affiliation{School of Physical Science and Technology, Ningbo University, Ningbo 315211, China}


\author{Yin-Zhe Ma}
\thanks{Corresponding author: Y.-Z. Ma, \url{mayinzhe@sun.ac.za}}
\affiliation{Department of Physics, Stellenbosch University, Matieland 7602, South Africa}


\begin{abstract}
The Generalized Needlet Internal Linear Combination (GNILC) method is a non-parametric component separation algorithm to remove the foreground contamination of the 21-cm intensity mapping data. In this work, we perform the Discrete Cosine Transform (DCT) along the frequency axis in the expanded GNILC framework (denoted eGNILC) which helps reduce the power loss in low multipoles, and further demonstrate its performance. We also calculate the eGNILC bias to modify the criterion for determining the degrees of freedom of the foreground ({\it dof}), and embed the Robust Principal Component Analysis (RPCA) in mixing matrix computation to obtain a blind component separation method. We find that the eGNILC bias is related to the averaged domain size and the {\it dof} of the foreground but not the underlying 21-cm signal. In case of no beam effect, the eGNILC bias is negligible for simple power law foregrounds outside the Galactic plane. We also examine the eGNILC performance in the SKA-MID (SKA Phase-I in mid-frequency) and BINGO (Baryon Acoustic Oscillations from Integrated Neutral Gas Observations) simulations. We show that if the adjacent frequency channels are not highly correlated, eGNILC can recover the underlying 21-cm signal with good accuracy. With the varying Airy-disk beam applied to both SKA-MID and BINGO, the power spectra of 21-cm can be effectively recovered at the multipoles $\ell \in [20, 250]$ and $[20, 300]$ respectively. With no instrumental noise, the SKA-MID exhibits $\lesssim 20\%$ power loss and BINGO exhibits $\sim 10\%$ power loss. The varying Airy-disk beam only causes significant errors at large multipoles. 
\end{abstract}



\section{introduction}
\label{sec:intro}
Intensity mapping (IM) of the neutral hydrogen 21-cm line has been studied as a promising technique to probe the large-scale structure (LSS) of the Universe~\citep{Madau:1996cs,Battye:2004re,Masui2013,Santos:2004ju,Bull:2014rha,Santos:2015gra,Bigot-Sazy2015,Bigot-Sazy2016,LiMa2017,Harper:2017gln,Bacon:2018dui,Xu2018,Anderson2018}. Instead of measuring the emission from individual galaxies, IM measures integrated emission from a voxel of many galaxies in which the total signal can be detected by radio telescopes. Therefore, the integrated $\HI$ flux density in each pixel of sky maps contains multiple galaxies' contributions to the signal.

Many experiments have been conducted or proposed to detect the $\HI$ power spectrum from auto- or cross-correlations. The techniques are mainly two kinds, the interferometer array and the single-dish (SD) measurement. The interferometer array measures the Fourier transform of the sky brightness in $uv$-plane (e.g. visibilities), which directly corresponds to the power spectra of the 21-cm signal. These include the Low-Frequency Array (LOFAR;~\citealt{Gehlot2019,Mertens2020}), the Precision Array for Probing the Epoch of Reionization (PAPER;~\citealt{Parsons2010,Gogo2022}), the Murchison Widefield Array (MWA;~\citealt{Trott2020}), and the Hydrogen Epoch Reionization Array (HERA;~\citealt{DeBoer2017,HERA2022a,HERA2022b,HERA2023}). The single-dish experiment, on the other hand, measures the specific flux of the 21-cm signal by direct imaging, which has achieved several detections in recent years via cross-correlation, including the HIPASS cross-correlation with 6dFGS galaxies~\citep{Pen:2008fw}, Green Bank Telescope (GBT) cross-correlation with DEEP2 galaxy survey~\citep{Chang2010} and with WiggleZ survey~\citep{Masui2013}, Parkes telescope cross-correlation with 2dFGS galaxies~\citep{Anderson2018}, and stacking of Parkes 21-cm fields with galaxy halos~\citep{Tramonte2020} and filaments~\citep{Tramonte2019}. More recently, \citet{Cunnington:2022uzo} utilized $10.5$-hour observations of MeerKAT ($64$-dishes) drifted scan over $\sim 200\deg^2$ and cross-correlated it with WiggleZ galaxies, and achieved $7.7\sigma$ detection of the cross-correlation signal. In the near future, Five-hundred-meter Aperture Spherical radio Telescope (FAST,~\citealt{Nan2011,Bigot-Sazy2016}), BINGO~\citep{Battye2012,Battye:2012tg,Dickinson2014,Battye2016,Wuensche:2019cdv,Yohana2019,Abdalla2022,Wuensche2022}, and SKA-MID~\citep{Bacon:2018dui} will also utilize the SD IM technique to measure the 21-cm brightness of the sky and probe the LSS.

In practice, the detection of $\HI$ IM signal is a great challenge because the foreground emission is at least $4$-orders of magnitude higher than the underlying signal. Fortunately, the spectral shape of the $\HI$ signal is highly fluctuated, which is in principle, separable from the otherwise smooth continuum foreground emission. Unlike the interferometric experiments, the single-dish experiment has to remove foreground contaminants to achieve the auto- or cross-spectrum of 21-cm fluctuations. Currently, there are two main techniques for foreground removal: parametric and non-parametric methods. Parametric methods use a (prior) model to describe the physical properties of the foregrounds. In contrast, the non-parametric methods try to minimize the use of prior knowledge of the foreground, but rather use the observed data for signal extraction. Some non-parametric methods have been developed to extract 21-cm foregrounds, such as the principal component analysis (PCA;~\citealt{Masui:2012zc}, \citealt{Alonso:2014dhk}, \citealt{Yohana:2021xck}, \citealt{Cunnington:2022uzo}), the independent component analysis (ICA;~\citealt{Alonso:2014dhk}), the fast independent component analysis (FASTICA;~\citealt{Maino:2001vz}, \citealt{Wolz:2013wna}), etc. 

In this paper, we study and expand the Generalized Needlet Internal Linear Combination (GNILC) method which was first proposed in ~\citet{Remazeilles:2011ze} and applied to the 21-cm IM in~\citet{Olivari:2015tka}, to extend the generalization and robustness of the GNILC method. We denote the expanded GNILC as ``eGNILC'' in this paper. We will incorporate the robust principal component analysis (RPCA) into the framework of GNILC, which becomes independent of the detailed knowledge of the underlying 21-cm signal. We name it as ``RPCA-embeded'' in the subsequent analysis of this paper. We will also evaluate the method in real-data scenarios and forecast its performance for BINGO and SKA-MID experiments.

This paper is organized as follows. In Sec.~\ref{sec:sky}, we describe the models for the $\HI$ signal, foreground components, and instrumental noise. In Sec.~\ref{sec:fgc}, we present the mathematical framework of the eGNILC and make an extensive discussion of its function on foreground cleaning.
Then in Sec.~\ref{sec:demon}, we demonstrate the performance of the foreground removal methods, and apply it to the simulated data of the BINGO and SKA-MID configurations in~Sec.~\ref{sec:gnilc_impl}. The conclusion is presented in Sec.~\ref{sec:conclusions}. 

\section{sky maps}
\label{sec:sky}
\begin{figure*}[htbp!]
  \centering
  \includegraphics[width=0.95\textwidth,keepaspectratio]{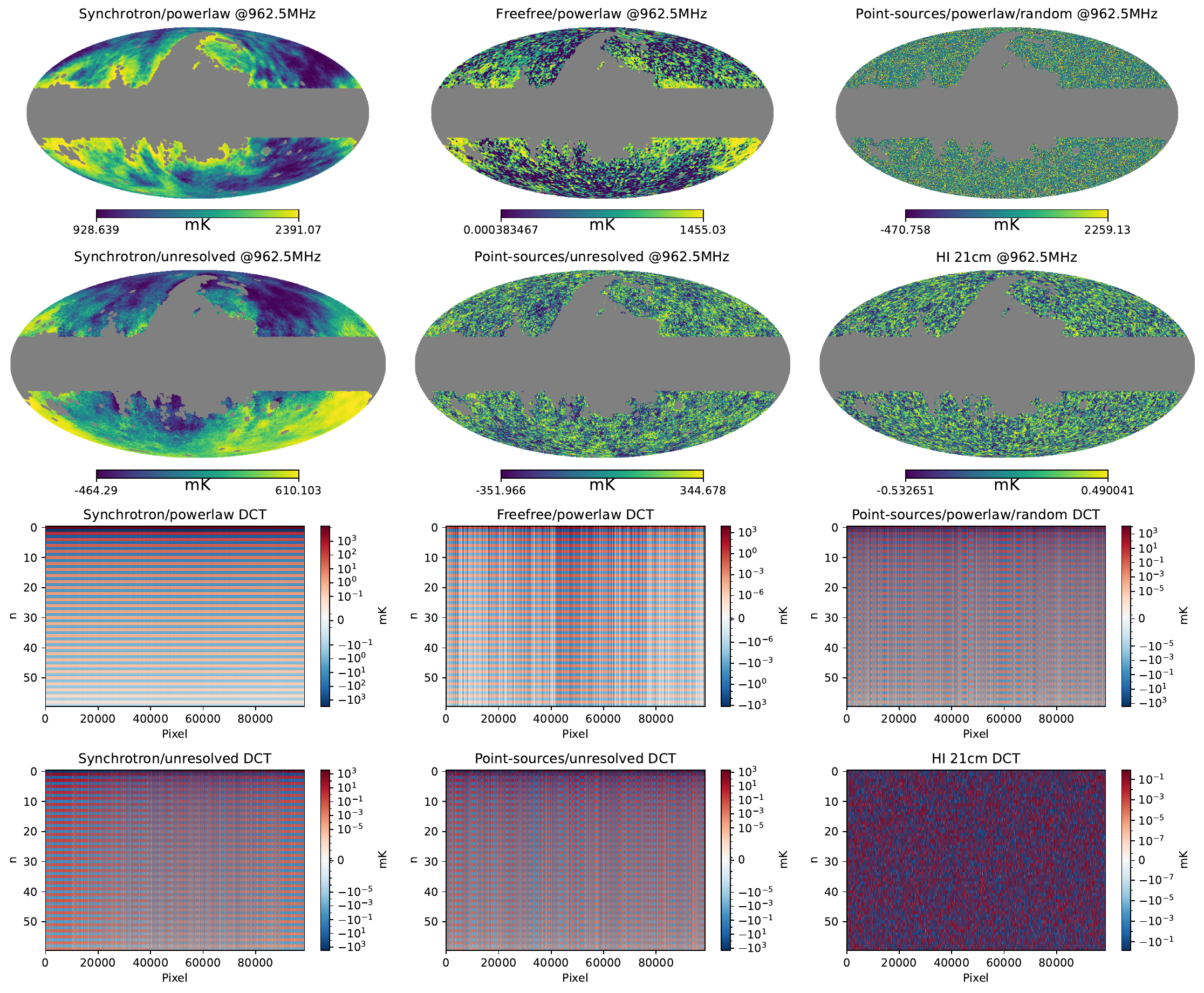}
  \caption{Sky maps of the synchrotron emission, free-free radiation, point sources, and 21-cm signal of $\HI$ at the frequency $962.5\,\text{MHz}$. All maps are normalized with histogram equalized color mapping. The discrete cosine transforms of $60$ channels ($962.5$-$1257.5\,\text{MHz}$) for individual catalogues are plotted in the square panels, where only unmasked pixels are used. The horizontal ticks label the pixel index, the vertical ticks label the DCT modes, and the color bar represents the DCT coefficients of maps.}
  \label{fig:sim_masked_maps} 
\end{figure*}

We simulate both $\HI$ signals and foregrounds in the interested frequency range.
Figure~\ref{fig:sim_masked_maps} shows the simulated synchrotron emission, free-free radiation, point sources, and the $\HI$ 21-cm signal in the upper (elliptical) panels.
The model of each component will be introduced in the following subsections. Throughout this paper, the resolution of all HEALPix~\citep{2005ApJ...622..759G,Zonca2019} maps is set to $n_\mathrm{side}=128$. In Fig.~\ref{fig:sim_masked_maps} (lower, square, panels), we show the discrete cosine transform (DCT) of the $60$ frequency channels spanning from $962.5\,\text{MHz}$-$1257.5\,\text{MHz}$ (centre frequencies). 
To see the smoothness of each component along the frequency, we use the Python \textsc{Scipy} package to conduct the DCT, which decomposes the data series at each pixel with cosine functions. We use the type-II DCT in the analysis, which is
\begin{equation}
    y_k = 2 f \sum_{n=1}^{N-1} x_n\cos\left( \frac{\pi k(2n+1)}{2N} \right) \,
\end{equation}
where the scaling factor is
\begin{equation}
    f = \left\{
        \begin{array}{cc}
             \sqrt{1/4N} & \text{if~} k=0\,, \\
             \sqrt{1/2N} & \text{otherwise}\,, 
        \end{array}
    \right.
\end{equation}
where $N$ is the number of frequency channels, and the index $n$ rum over frequencies.

\subsection{$\HI$ 21-cm intensity map}

The post-reionization $\HI$ 21-cm signal traces the distribution of hydrogen.
Ignoring the perturbations, the background brightness temperature of 21-cm is proportional to the number density of neutral hydrogen $\bar{n}_\mathrm{\HI}$~\citep{Hall:2012wd}
\begin{equation}
\bar{T}_\mathrm{b}(z) = \frac{3(h_{\rm p} c)^3 \bar{n}_\mathrm{\HI}A_\mathrm{10}}{32\pi k_\mathrm{B} E^2_\mathrm{21}(1+z)H(z)}\,,
\label{eq:HI21cm-temp}
\end{equation}
where $h_\mathrm{p}$ is Planck's constant, $c$ is the speed of light in vacuum, $k_{\rm B}$ is Boltzmann's constant, $A_\mathrm{10}=2.869\times 10^{-15}\,{\rm s}^{-1}$ is the spontaneous emission coefficient, and $E_\mathrm{21}=5.87\,\mu{\rm eV}$ is the rest-frame energy of the 21-cm photon.

The perturbed 21-cm temperature is straightforward if we neglect the peculiar velocities and the Sachs-Wolfe effect~\citep{Hall:2012wd,Battye:2012tg}
\begin{equation}
\delta T_\mathrm{b}(z, \vec{r}(z)) = \bar{T}_\mathrm{b}(z)\delta_\mathrm{\HI}(z, \vec{r}(z))\,,
\label{eq:HI21cm-delta}
\end{equation}
where $\delta_\mathrm{\HI}(z, \vec{r}(z))$ is the fluctuation of local $\HI$ over-density and $\vec{r}(z)$ is the comoving position of 21-cm sources.
Making a Fourier transform and expressing $\delta_\mathrm{\HI}(z, \vec{r}(z))$ in the spherical Bessel function basis, we have
\begin{eqnarray}
\hspace*{-5mm}
\delta T_\mathrm{b}(z, \vec{r}(z)) &=& \bar{T}_\mathrm{b}(z) \int \frac{{\rm d}^{3}\vec{k}}{(2\pi)^3}\tilde{\delta}_\HI (\vec{k}, z) e^{i \vec{r}(z) \cdot \vec{k}} \nonumber \\
&=& 4\pi \bar{T}_\mathrm{b}(z) \sum_{\ell m} i^\ell \int \frac{{\rm d}^{3}\vec{k}}{(2\pi)^3} \tilde{\delta}_\HI (\vec{k}, z) \nonumber \\
&& \hspace*{15mm} \times j_\ell(k r(z)) Y_{\ell m}^\star (\hat{k}) Y_{\ell m}(\hat{r})\,, \label{eq:delta_Tb}
\end{eqnarray}
where we used the decomposition of a plane wave $e^{i \vec{k} \cdot \vec{r}} = 4 \pi \sum_{\ell m} i^\ell \, j_\ell(k r(z)) \, Y_{\ell m}^\star (\hat{k}) \, Y_{\ell m}(\hat{r})$. From Eq.~(\ref{eq:delta_Tb}) we obtain the corresponding harmonic coefficients of redshift-slice $z$
\begin{eqnarray}
a_{\ell m}(z) &=&  4\pi i^{\ell}\, \bar{T}_\mathrm{b}(z) \int \frac{{\rm d}^{3}\vec{k}}{(2\pi)^3} \tilde{\delta}_\HI (\vec{k}, z) \nonumber \\
& \times & j_\ell(k r(z)) Y_{\ell m}^\star (\hat{k}). \label{eq:alm}
\end{eqnarray}
We now calculate the angular power spectra of 21-cm fluctuations. Because the density contrast is translational and rotational invariant, we can define the $\HI$ power spectra as 
\begin{eqnarray}
&& \hspace*{-4mm} \langle \tilde{\delta}_\HI (\vec{k}, z)\tilde{\delta}^{\ast}_\HI (\vec{k^\prime}, z^\prime) \rangle \nonumber \\
&& \hspace*{8mm} = (2\pi)^3 \delta^{(3)}_{\mathrm{D}}(\vec{k} - \vec{k^\prime}) P_\HI (k, z) \nonumber \\
&& \hspace*{8mm} = (2\pi)^3 \delta^{(3)}_{\mathrm{D}}(\vec{k} - \vec{k^\prime}) b^2 P_\mathrm{m} (k) D(z) D(z^\prime)\,,
\end{eqnarray}
where in the last line, we utilized the fact that the $\HI$ signal traces the underlying matter density field on large scales after reionization. $\delta^{(3)}_{\mathrm{D}}$ is the 3-D Dirac delta function, $P_\HI$ and $P_\mathrm{m}$ are the $\HI$ and matter power spectra, $D(z)$ is the growth factor, and $b$ is the bias between the spatial distribution of the $\HI$ and the matter field. Then by ensemble-averaging the quadratic $a_{\ell m}$ (Eq.~(\ref{eq:alm})), we obtain the angular power spectrum of 21-cm fluctuations in redshift bins ($z,z^{\prime}$; see also~\citealt{Battye:2012tg})
\begin{eqnarray}
C_{\ell}(z, z^\prime) &=& \frac{2 b^2}{\pi} \bar{T}_\mathrm{b}(z) \bar{T}_\mathrm{b}(z^\prime)  D(z) D(z^\prime) \nonumber \\
&\times & \int \mathrm{d}k\,k^2  P_\mathrm{m}(k) j_\ell(k r(z)) j_\ell(k r(z^\prime))\,.
\end{eqnarray}

We use the publicly available code Cosmology in the Radio Band (CORA)\footnote{\url{https://github.com/radiocosmology/cora}} to simulate the $\HI$ 21-cm intensity map at low redshifts~\citep{Shaw:2013wza}.
The upper (elliptical) panels in Fig.~\ref{fig:sim_masked_maps} show the maps at $962.5\,\text{MHz}$ as an example.
The amplitude of the 21-cm fluctuation is around $0.1\,$mK, which is roughly $4$ order smaller than the foreground.
The DCT of 21-cm maps (bottom-right panel) shows correlations among frequencies are weak because the magnitude of each DCT mode is roughly at the same level. Therefore, a highly efficient foreground removal method and fast computational package are essential to extract the heavily contaminated signal.

Due to the limited sky coverage and Galactic contamination, observations are usually conducted in the partial sky, where the full-sky power is modulated with sky masks and mode-mixing effect. The ``pseudo-$C_{\ell}$'' scheme shows that the pseudo-power spectrum $\tilde{C}_\ell$ on a cut sky is related to the full-sky power spectrum $C_\ell$ via
\begin{eqnarray}
\langle \tilde{C}_\ell \rangle = \sum_{\ell^\prime} M_{\ell\ell^\prime} B^2_{\ell^\prime}\langle C_{\ell^\prime} \rangle , \label{eq:pcl}
\end{eqnarray}
where $M_{\ell\ell^\prime}$ is mode-mixing matrix, and $B_{\ell}$ takes care of the telescope beam and finite pixelization effect~\citep{Hivon:2001jp,Efstathiou2004,Osato:2019ynf}.
The mode-coupling matrix is
\begin{equation}
M_{\ell_1 \ell_2} = \frac{2\ell_2 + 1}{4 \pi} \sum_{\ell_3} (2\ell_3 +1) W_{\ell_3} \left(
\begin{array}{ccc}
\ell_1 &\ell_2 &\ell_3 \\
0 &0 &0
\end{array}
\right)^2 \,, \label{eq:Mell}
\end{equation}
where $W_\ell$ is the power spectrum of the mask
\begin{eqnarray}
W_{\ell}&= & \frac{1}{2\ell +1}\sum_{m}w_{\ell m}w^{\ast}_{\ell m} \nonumber \\  
w_{\ell m} &=&
\int {\rm d}^{2}\hat{\theta}\,w(\hat{\theta})Y^{\ast}_{\ell m}(\hat{\theta}),
\end{eqnarray}
and $w(\vec{\theta})$ is the mask function. The last term in Eq.~(\ref{eq:Mell}) is the Wigner-$3j$ symbol (Clebsch-Gordan coefficient;~\citealt{Abramowitz1972}).

\subsection{Foregrounds}
\label{subsec:foregrounds}
The Galactic synchrotron emission dominates the foregrounds at the frequency bands of 21-cm experiments. In addition, the free-free radio emission from ionized gas and extragalactic point sources are also much brighter than the underlying $\HI$ 21-cm signal. Because of the spectral smoothness of the foregrounds, many works have studied the scaling laws of the foregrounds in frequency direction to extrapolate the foreground at a given frequency from the template maps~\citep{Haslam1982,Platania:1997zn,MivilleDeschenes:2008hn,deOliveiraCosta:2008pb,Kogut2012,Remazeilles:2014mba}. However, these templates may not be adequate for including all possible foreground components. In this work, we also add unresolved point sources to fully examine the robustness of our foreground removal method. We present each of the foreground components below. The synchrotron emission and point sources are simulated with CORA code~\citep{Shaw2015} and the free-free radiation with PySM\footnote{PySM is a publicly available Python code generating full-sky simulations of Galactic foregrounds in both intensity and polarization, relevant to CMB experiments.}~\citep{Thorne:2016ifb}.

\subsubsection{Extrapolation of sky maps}
\label{subsec:extrapolation_sky}
\begin{itemize}
\item Synchrotron emission. The simulation of Galactic synchrotron emission is based on the all-sky $408\,\text{MHz}$ Haslam map~\citep{Haslam1982,Remazeilles:2014mba}.
New maps can be generated by scaling the brightness temperature map $T_\text{Has}$ at desired frequencies with a power-law relation
\begin{eqnarray}
T_{\text{syn}}(\hat{n}, \nu) &=& T_\text{Has}(\hat{n},408\,\mathrm{MHz}) \nonumber \\
& \times &
\left(
\frac{\nu}{408\,\mathrm{MHz}}
\right)^{\alpha(\hat{n})}\,,
\end{eqnarray}
where $\alpha(\hat{n})$ is the index per pixel determined by the slope of the logarithmic intensities at $408\,\text{MHz}$ and $1420\,\text{MHz}$
\begin{equation}
\alpha(\hat{n})=\frac{\log T_{1420}(\hat{n}) - \log T_{408}(\hat{n}) }
{\log{1420} -\log{408}}\,.
\end{equation}
The $408\,\text{MHz}$ and $1420\,\text{MHz}$ maps are generated by fitting the global sky model from surveys with different sky coverage and ranging from $10\,\text{MHz}$ to $94\,\text{GHz}$~\citep{deOliveiraCosta:2008pb}. (See the first row, first column map in Fig.~\ref{fig:sim_masked_maps} for a simulated synchrotron map. 

\item Free-free radiation. We use a nominal model to simulate the free-free emission with PySM code (first row, second column in Fig.~\ref{fig:sim_masked_maps}.) We first generate an intensity map at $\nu_0=30\,\text{GHz}$, and then scale brightness temperature with a spatially constant power law index $-2.14$~\citep{Draine2011}, which is supposed to be valid for the frequency down to $10\,\text{MHz}$~\citep{Thorne:2016ifb,Adam:2015wua}:
\begin{eqnarray}
T^{\text{ff}}_\text{b}(\hat{n})=T^{\text{ff}}_{{\rm b}0}(\hat{n}) \left(\frac{\nu}{\nu_0}\right)^{-2.14}\,.
\end{eqnarray}


\item Point sources. Because bright point sources are well measured and catalogued, point sources with flux $S>10\,\mathrm{Jy}$ are assumed to be subtracted from the maps easily when we obtain an IM volume. Therefore, we generate medium bright point sources ($100\,\mathrm{mJy}<S<10\,\mathrm{Jy}$) from the point source distribution model of~\citet{DiMatteo:2001gg}. We show an example of this generation in the first row, the third column of Fig.~\ref{fig:sim_masked_maps}. Because the discrete point sources map is not band-limited, spherical transforms of the map may be inaccurate such as the negative pixel values shown in Fig.~\ref{fig:sim_masked_maps}. We will also introduce the unresolved foregrounds below to take care of this effect and other dim point sources. 
\end{itemize}

\subsubsection{Unresolved random sources}
\label{subsec:unresolved_sky}
In addition to the power-law foreground maps, we also incorporate the extra unresolved random sources imported in the CORA code to fully study the effect of foreground removal of our computational method. The angular power spectrum of unresolved point sources can be parametrized as~(\citealt{Shaw:2013wza}; see also~\citealt{Santos:2004ju})
\begin{eqnarray}
C_{\ell}(\nu, \nu^\prime) = A \left(\frac{\ell}{\ell_0}\right)^{-\alpha}
\left(\frac{\nu \nu^\prime}{\nu_0^2}\right)^{-\beta}\!
e^{-\left[\ln^2(\nu/\nu^\prime)\right]/2\zeta^{2}}\!,
\label{eq:aps_random_foreground}
\end{eqnarray}
where $\ell_0 = 100$ and $\nu_0 = 408\,{\rm MHz}$. The values of the parameters are assumed to be
\begin{eqnarray}
\left[A\,({\rm mK}^{2}), \alpha, \beta, \zeta \right]_{\rm syn} &=& \left[6.6\times 10^{3},\, 2.80, \, 2.8, \, 4.0 \, \right], \nonumber \\
\left[A\,({\rm mK}^{2}), \alpha, \beta, \zeta \right]_{\rm PS} &=& \left[35.5,\, 2.10, \, 1.1, \, 1.0 \, \right],
\end{eqnarray}
for synchrotron and point sources emissions respectively~\citep{Shaw:2013wza}.
Using Eq.~(\ref{eq:aps_random_foreground}) and assuming Gaussian distribution, we simulate the unresolved synchrotron emission and point sources.
They are shown in the second row, first and second columns in Fig.~\ref{fig:sim_masked_maps} respectively. 

Equation~\eqref{eq:aps_random_foreground} shows that the unresolved random foreground depends on its frequency and spatial distribution. 
The discrete cosine transform reflects important spectral and spatial features of the $\HI$ signal and foregrounds.
For the power-law foreground contamination, the magnitudes of DCT coefficients decrease rapidly as the index $n$ increases, which suggests that these foregrounds can mostly be determined by a few principal components. In other words, this feature makes the foreground easier to remove.

The spatial patterns are different in the DCT of power-law foregrounds because the synchrotron map is spatially smooth while temperature fluctuations are more dramatic in the free-free and point-source maps. The unresolved maps of both synchrotron and point sources are stochastic ``noises".
Because the angular power spectrum of the unresolved synchrotron decreases faster ($\alpha$ is larger) than that of the unresolved point-source map, the map of the unresolved synchrotron foreground looks smoother than the unresolved point-source map. The horizontal stripes of the DCT images also suggest such spatial smoothness, while descending mode coefficients reflect the correlation of the $\nu$ and $\nu^\prime$ angular power spectrum. An outstanding feature of the $\HI$ signal is that the mode coefficients do not decay visibly.

\subsubsection{Instrumental noise}
\label{subsubsec:noise}
The most common instrumental noise is the Gaussian distributed thermal noise.
The {\it rms} of pixel noise is related to the system temperature, integration time, and frequency resolution~\citep{Bigot-Sazy2015,Olivari:2015tka}
\begin{equation}
\sigma_\mathrm{pix} = \frac{T_{\mathrm{sys}}}{\sqrt{t_{\mathrm{pix}}\delta \nu}}\,,
\end{equation}
where $\delta \nu$ is the width of each frequency channel, $T_{\mathrm{sys}}$ is the system temperature, and $t_{\mathrm{pix}}$ is the total integration time per pixel. We can simulate thermal noise according to different instrumental specifications in Table~\ref{tab:instru_param}.

We notice that the $1/f$ noise also plays an important role in the radio antenna system. \cite{Harper:2017gln} studied the impacts of $1/f$ noise on $\HI$ IM experiments by simulations with the following power-law model of the power spectral density (PSD),
\begin{eqnarray}
\mathrm{PSD}(f,\omega) = \frac{T^2_\mathrm{sys}}{\delta \nu}\left[ 
1 + C(\beta, N_\nu)\frac{\left(f_{\rm k}/f\right)^\alpha}{(\omega\Delta\nu)^\frac{1-\beta}{\beta}}\right] \,,
\end{eqnarray}
where $f_{\rm k}$ is the knee frequency, $\alpha$ is the spectral index, $\omega$ is the inverse spectroscopic frequency wavenumber, $\Delta\nu$ is the total receiver bandwidth, $\beta$ is used to parametrize the spectral index of the PSD\footnote{We remind the reader not to confuse the meaning of $\alpha$ and $\beta$ here with the parameters defined in unresolved foreground sky (Sec.~\ref{subsec:unresolved_sky}).}, and $C(\beta, N_\nu)$ is a normalization factor. Because it is spectrally uncorrelated, pure thermal noise is irremovable but reducible with more integration time, while the $1/f$ noise is removable for data with high correlations between frequency channels. \cite{Olivari:2015tka} explored the instrumental effects by adding thermal noise in the simulation. They presented the residual maps for the cases with no thermal noise and with thermal noise amplitudes equal to $0.05\mathrm{mK}$ and $0.08\mathrm{mK}$. It is concluded that the larger the thermal noise amplitude is, the less accurate the reconstruction of the \HI power spectrum on small angular scales is. They also suggest a solution in the case where there is a good estimate of the thermal noise power spectrum. The \HI +noise covariance matrix is used as the prior instead of the \HI covariance matrix, so that we recover the \HI signal plus thermal noise. We may correct for the thermal noise by using the estimate of the thermal noise power spectrum. For effects of the $1/f$ noise, readers may refer to \citet{Harper:2017gln} and~\citet{Yohana:2021xck}. Therefore, we primarily focus on removing the foreground contamination, and will not discuss the impacts of thermal noise or $1/f$ instrumental noise in the subsequent discussion.

\begin{table*}[!htbp]
\renewcommand\arraystretch{1.5}
\begin{center}
\caption{Experimental specifications of simulations for the demonstration, the SKA-MID~\citep{Bacon:2018dui, Wang:2020lkn} and the BINGO~\citep{Battye2012,Battye:2012tg,Dickinson2014,Battye2016,Wuensche:2019cdv,Yohana2019,Abdalla2022,Wuensche2022} configurations, where the subscript FWHM stands for the full width of half maximum. Without loss of generality, we vary the channel width for convenience.}
\begin{ruledtabular}
\begin{tabular}{ccccc}
\multirow{2}{*}{Parameters} & \multicolumn{3}{c}{Values} &\multirow{2}{*}{Comments} \\ \cline{2-4}
\multirow{2}{*}{}           &Demonstration &SKA-MID &BINGO &\multirow{2}{*}{}\\
\hline
$[z_{\text{min}}, z_{\text{max}}]$                & $[0.13, 0.48]$ & $[0.33, 0.46]$  & $[0.13, 0.48]$    & Redshift range\\
$[\nu_{\text{min}}, \nu_{\text{max}}]\,(\text{MHz})$ & $[960, 1260]$  & $[970, 1070]$ & $[960, 1260]$  & Band\,width \\
$\delta \nu\,(\rm MHz)$                             & $15$/$5$  & $0.2$  & $5$           & Channel width \\
$\Omega_{\text{sur}}\,(\text{deg}^2)$                & Full sky       & $20000$   & $5000$      & Sky coverage \\
$\theta_{\text{FWHM}}\,(\text{arcmin})$              & N/A            & $80~@970\rm MHz$ & $40~@960\rm MHz$ & Beam\,width \\

\end{tabular}
\end{ruledtabular}
\label{tab:instru_param}
\end{center}
\end{table*}

\section{Foreground Removal Method}
\label{sec:fgc}

The GNILC method was first applied to 21-cm IM data by~\citet{Olivari:2015tka}. We will utilize this framework in general, but replace the theoretical cross-correlation matrix of the 21-cm signal with the one estimated by the RPCA method in one case. We will also elaborate on the refinement of the GNILC method and highlight the differences and improvements.

\subsection{GNILC}
\label{subsec:GNILC}
We denote the observed sky at the pixel $p$ and frequency $i$ as $x_i(p)$, which sums the fluctuation of $\HI$ 21-cm emission $s_i(p)$ and the foreground components $f_i(p)$. We remind the reader that $p$ does not necessarily mark the pixels in the real space, and it could be the index of the needlet coefficient in the GNILC. 
Similarly, the index $i$ may label any variant of frequency, such as the mode index of DCT shown in Fig.~\ref{fig:sim_masked_maps}. We may write the dependence of the frequency index in the vector form as
\begin{equation}
\boldsymbol{x}(p)=\boldsymbol{s}(p) + \vf(p) \,, 
\label{eq:x_s_f}
\end{equation}
where $\boldsymbol{x}(p)$, $\boldsymbol{s}(p)$, and $\boldsymbol{f}(p)$ are $n_\text{ch}$-dimensional column vectors, and $n_\text{ch}$ is the number of frequency channels. The signal $\boldsymbol{s}$ is, of course, independent of the foreground $\boldsymbol{f}$. The bold characters denote the vectors (lowercase) or matrices (uppercase) defined in the frequency basis. From Eq.~(\ref{eq:x_s_f}), the covariance matrix at pixel $p$ is
\begin{eqnarray}
\bR_{\bx}(p)=\bR_{\bs}(p) + \boldsymbol{R}_{\vf}(p) \,, 
\label{eq:R_x=HI+n}
\end{eqnarray}
where the three $\boldsymbol{R}$\,s are the covariance matrices of $\boldsymbol{x}$, $\boldsymbol{s}$, and $\boldsymbol{f}$ respectively. Below we suppress the variable $p$ for brevity.


Because of the smoothness of foreground on frequency direction, the number of independent modes of foregrounds $m$ is much less than the frequency channel number $n_{\rm ch}$, i.e. $m \ll n_{\rm ch}$. As a consequence, the recovered $\HI$ signal $\hat{\boldsymbol{s}}(p)$ may only have $n_\text{ch}-m$ degrees of freedom ({\it dof}\,), regardless of its real {\it dof}. For the case where thermal noise is presented, the noise still exists in the $\HI$ 21-cm signal after the foreground removal procedure because only the frequency-correlated contaminants are removed. Because of $dof=n_\text{ch}-m$, the estimated 21-cm signal $\hat{\boldsymbol{s}}$ is a linear combination of independent templates $\boldsymbol{t}$ as
\begin{eqnarray}
\hat{\boldsymbol{s}}=\boldsymbol{S}\boldsymbol{t}\,,
\end{eqnarray}
where $\boldsymbol{S}$ is an $n_\text{ch}\times(n_\text{ch}-m)$ mixing matrix.
Then the covariance matrix of $\hat{\boldsymbol{s}}$ is the transformation of the full-rank matrix $\boldsymbol{R}_\text{t}$:
\begin{eqnarray}
\boldsymbol{R}_{\hat{\bs}}=\boldsymbol{S}\boldsymbol{R}_{\vt}\boldsymbol{S}^T\,,
\end{eqnarray}
where $\boldsymbol{R}_\text{t}$ is the covariance matrix of $\boldsymbol{t}$. One should realize that the mixing matrix ($\boldsymbol{S}$) and template set ($\boldsymbol{t}$) are not unique, because, for any given set of templates, a new set of templates ($\boldsymbol{t}$') can be generated with an invertible linear transformation
\begin{eqnarray}
\boldsymbol{t}^\prime&=&\boldsymbol{V}^{-1}\boldsymbol{t}\,, \\
\boldsymbol{R}_{\hat{\bs}}&=&\boldsymbol{S}\boldsymbol{V}\boldsymbol{R}_{\vt^\prime}
\boldsymbol{V}^T\boldsymbol{S}^{T}\,.
\label{eq:R_s_t^prime}
\end{eqnarray}

For a given mixing matrix, we want to find an ``ILC weighing matrix'' $\boldsymbol{W}$ 
\begin{eqnarray}
\hat{\boldsymbol{s}}=\boldsymbol{W}\boldsymbol{x}\,
\label{eq:extract_s}
\end{eqnarray}
to extract the $\HI$ signal $\hat{\boldsymbol{s}}$ (which is generally believed to deviate from $\boldsymbol{s}$) such that the total variance is minimized, and the 21-cm signal is unbiased
\begin{eqnarray}
\min\left[\text{Tr}(\boldsymbol{W}\bR_{\bx}\boldsymbol{W}^T)\right],~{\rm s.t.}~\boldsymbol{W}\boldsymbol{S}=\boldsymbol{S}\,.
\end{eqnarray}
The solution of this conditional minimization is analogous to the famous ``map-making'' equation~\citep{Tegmark1996,Dodelson2020}
\begin{eqnarray}
\boldsymbol{W}=\boldsymbol{S}(\boldsymbol{S}^T\bR_{\bx}^{-1}\boldsymbol{S})^{-1}
\boldsymbol{S}^T\bR_{\bx}^{-1}\,.
\label{eq:mixing}
\end{eqnarray}

Therefore, with the weighing matrix (Eq.~(\ref{eq:mixing})), we can implement the needlet-space ILC with the following steps:
\begin{itemize}
\item {\it Analysis}. First, the real map at a particular frequency is needlet-transformed while the needlets have a finite spectral support adjustable at will\footnote{In other words, each needlet map is band-limited according to the choice of window function.} and have a good spatial localization.
\begin{eqnarray}
\{x(p)\}_{p\in \mathcal{P}}&\xrightarrow{\text{SHT}}&
\{a_{\ell m}\}\xrightarrow{\times}\{b^{(j)}_\ell a_{\ell m}\}\nonumber\\
&\xrightarrow{\text{SHT}^{-1}}&
\{x^{(j)}(p^\prime)\}_{p^\prime\in \mathcal{P}^{(j)}}\, ,
\nonumber
\end{eqnarray}
where $x(p)$ is the data at each pixel $p$ on the map $\mathcal{P}$, $\text{SHT}$ is the acronym of Spherical Harmonics Transform, $\text{SHT}^{-1}$ is the inverse transform, ${b^{(j)}_\ell}$ is the window family that defines the needlet bases, and $\{x^{(j)}(p^\prime)\}_{p^\prime\in \mathcal{P}^{(j)}}$ is the projected map corresponding to each needlet base. The 21-cm signal will be extracted from individual needlet map and then synthesized with the following steps.

\item {\it Signal extraction}. As illustrated above, we obtained the expected signal extracted from the decomposed data in the needlet space with Eq.~(\ref{eq:extract_s}).
As the weighing matrix is yet to be solved, Sec.~\ref{subsubsec:mixing_matrix} will introduce the covariance ($\bR_{\bx}$) and mixing matrics ($\boldsymbol{S}$), which are crucial for Eqs.~(\ref{eq:mixing}) and (\ref{eq:extract_s}) to extract the underlying signal.

\item {\it Synthesis}. We finally synthesize the extracted signal from each needlet base, which is the reversal of analysis.
\begin{eqnarray}
\{\hat{s}^{(j)}(p^\prime)\}_{p^\prime\in \mathcal{P}^{(j)}}
&\xrightarrow{\text{SHT}}&\{b^{(j)}_\ell a_{\ell m}\}
\xrightarrow{\times}\{{\tilde{b}}_\ell^{(j)}b^{(j)}_\ell a_{\ell m}\}\nonumber\\
&\xrightarrow{\text{SHT}^{-1}}&\{\Psi^{(j)}\hat{s}(p)\}_{p\in \mathcal{P}}\, , \nonumber
\end{eqnarray}
where $\{\tilde{b}^{(j)}_\ell\}$ is the synthesis window family that ensures the reconstruction unbias condition $\sum_{j\in\mathcal{J}}\tilde{b}^{(j)}_\ell b^{(j)}_\ell = 1$, and $\Psi^{(j)}$ is the resultant smooth operator at needlet mode $j\in \mathcal{J}$. The symbol $\hat{ }$ on $s$ refers to the estimate of $s$.
At each frequency, we get the expected 21-cm map by summing up all needlet modes $\hat{s}(p) = \sum_{j\in\mathcal{J}} \Psi^{(j)}\hat{s}^{(j)}$.
\end{itemize}

\subsubsection{Mixing Matrix}
\label{subsubsec:mixing_matrix}

Before determining the mixing matrix $\boldsymbol{S}$, we need to calculate covariance matrix of the signal.
In general, the sky maps observed at $n_\text{ch}$ frequency channels ($\boldsymbol{\Theta}(p)$), can be decomposed with spherical harmonics
\begin{eqnarray}
\boldsymbol{\Theta}(p)=\sum_{\ell m} \boldsymbol{a}_{\ell m} Y_{\ell m}(p)
=\sum_{j\in\mathcal{J}}\boldsymbol{\Theta}^{(j)}(p)\,,
\end{eqnarray}
where $Y_{\ell m}(p)$ is the spherical harmonic function, and the vector $\boldsymbol{\Theta}(p)$ represents the $p^{\mathrm{th}}$ pixel values on maps at all frequency channels. $\boldsymbol{\Theta}^{(j)}(p)=\sum_{\ell m} b^{(j)}_\ell \boldsymbol{a}_{\ell m} Y_{\ell m}(p)$ is the $j^{\rm th}$ needlet component. If the statistical isotropy holds, the covariance matrix can be calculated as
\begin{eqnarray}
\boldsymbol{R}^{(j)}&=&\langle\boldsymbol{\Theta}^{(j)}\boldsymbol{\Theta}^{(j)\, T}\rangle \nonumber \\
&=& \langle\boldsymbol{\Theta}^{(j)}\boldsymbol{\Theta}^{(j)\, \dagger}\rangle
\nonumber\\
&=&\sum_{\ell, m, m^\prime} \langle ({b}^{(j)}_\ell \boldsymbol{a}_{\ell m}) ({b}^{(j)}_\ell \boldsymbol{a}_{\ell m^\prime}^\dagger) \rangle Y_{\ell m} Y_{\ell m^\prime}^\ast
\nonumber\\
&=&\sum_{\ell, m, m^\prime} \delta_{m m^\prime} {B}^{(j)}_\ell \boldsymbol{C}_\ell Y_{\ell m} Y_{\ell m^\prime}^\ast
\nonumber \\
&=& \sum_{\ell} {B}^{(j)}_\ell \boldsymbol{C}_\ell \sum_{m} Y_{\ell m} Y_{\ell m}^\ast \nonumber\\
&=&\sum_{\ell}\frac{2\ell+1}{4\pi} {B}^{(j)}_\ell \boldsymbol{C}_\ell\,,
\label{eq:R_ell_p}
\end{eqnarray}
where the covariance matrix is pixel-independent, ${B}^{(j)}_\ell = {b}^{(j)}_\ell {b}^{(j)}_\ell$, and the matrix $\boldsymbol{C}_\ell$ consists of $n_\text{ch}\times n_\text{ch}$ cross-power spectra at $\ell$.

However, foregrounds are not isotropic. Instead of Eq.\,\eqref{eq:R_ell_p}, we calculate the covariance matrix of each pixel from the sample average, because maps are smooth at a specific scale for a given choice of needlet bases. For each needlet mode $j$, the covariance matrix at pixel $p$ is an average of the domain $\mathcal{D}$ around $p$. The effective independent pixel number is often smaller than the true pixel number in the domain because nearby pixels are correlated, especially for the needlet coefficient maps that are filtered with spectral window functions. The domain $\mathcal{D}$ can be defined by convolving maps with a symmetric Gaussian window in the real space so that the average is easily done in the harmonic domain. The size of $\mathcal{D}$ is related to the smoothing scale of the $j^{\rm th}$ needlet, but the GNILC is insensitive to the detailed form of $\mathcal{D}$. We will use this average as the estimate of covariance in our analysis. However, there is a caveat to this approach: the limited number of independent samples may cause (artificial) anti-correlation between the signal and contaminants, which leads to a power loss in the recovered signal~\citep{Delabrouille:2008qd,Olivari:2015tka}. We will derive this bias in Sec.~\ref{subsubsec:dofm}.

The spectral window function determines how the needlet map is localized in both the spatial and spherical harmonics domains (refer to Appendix~\ref{sec:app_needlet}). Thus, the choice of the parameter $B$ should be in line with the spatial and spectral features of the 21-cm signal and foregrounds. 
In addition to the dependence on the spectral window function of the needlet space, the determination of $\mathcal{D}$ is also affected by the foregrounds, masks, and telescope beams. We will show these effects in Sections\,\ref{sec:demon} and Sec.\,\ref{sec:gnilc_impl}. In our analysis, we use the effective  $\theta_\mathrm{FWHM}$ to quantify the needlet, beam, pixel size, and sky coverage in order to determine the domain $\mathcal{D}$ and $N_p$ - the effective number of independent pixels in $\mathcal{D}$. First of all, bear in mind that we want a large $N_p$ to obtain an accurate estimate of the covariance matrix and reduce the bias (see Sec.~\ref{subsubsec:dofm}). The independent size is limited by the map pixel size, the telescope beam, and the needlet size, which should be smaller than those three sizes. On the other hand, the maximum $\mathcal{D}$ is limited by the sky coverage (or mask), because if the $\mathcal{D}$ is significantly larger than the sky coverage, the data is averaged with the null field. In practice, the data and the null field are indeed blended together in the needlet domain, to a degree.

We can now calculate the mixing matrix $\boldsymbol{S}$. In the following, we will frequently use the square root of the real symmetric, positive-definite matrix $\bR_{\bs}$, which we define via its eigendecomposition, i.e., 
\begin{eqnarray}
\bR_{\bs}=\bQ\bLambda\bQ^\mathrm{T}  \Rightarrow \bR_{\bs}^{1/2}=\bQ\bLambda^{1/2}\bQ^\mathrm{T}\,. 
\end{eqnarray}
By using Eq.~\eqref{eq:R_x=HI+n}, we have
\begin{eqnarray}
\bR_{\bs}^{-1/2}\bR_{\bx}\bR_{\bs}^{-1/2}=\bR_{\bs}^{-1/2}\bR_{\vf}\bR_{\bs}^{-1/2} + \bI\,, \label{eq:LxL1}
\end{eqnarray}
where the superscript $(j)$ is omitted for simplicity. Note that any covariance matrix is a positive semidefinite matrix, and the signal is nonzero. Thus we can diagonalize the left-hand side of Eq.~(\ref{eq:LxL1}) as
\begin{eqnarray}
&& \bR_{\bs}^{-1/2}\bR_{\bx}\bR_{\bs}^{-1/2} \nonumber 
\\
&=&
[\bU_{\bs}\, \bU_{\vf}]
\left[
\begin{array}{cc}
\bI_{\{n_\mathrm{ch}-m\}} & \boldsymbol{0} \\
\boldsymbol{0} & \bD_{\vf} \\
\end{array}
\right]
\left[
\begin{array}{c}
\bU^T_{\bs}\\
\bU^T_{\vf}
\end{array}
\right] \nonumber \\ 
&=& \bU_{\vf} \bD_{\vf}\bU^T_{\vf} +
\bU_{\bs} \bU_{\bs}^{T} \,,
\label{eq:eigen}
\end{eqnarray}
where $\bU_{\bs}$ is an $n_\text{ch}\times (n_\text{ch}-m)$ sub-matrix, $\bI_{\{n_\mathrm{ch}-m\}}$ is an $(n_\text{ch}-m)\times (n_\text{ch}-m)$ unit matrix, and $\bD_{\vf}=\text{diag}(\lambda_1+1, \cdots, \lambda_m +1)$ ($\lambda_i \gg 1$ are foreground modes). $\bU_\text{f}$ is an $n_\text{ch} \times m$ sub-matrix, and $\bU_{\vf}\bU_{\vf}^{T}+\bU_{\bs}\bU_{\bs}^{T}=\bI$.
Then, the following expression gives an approximated foreground covariance matrix
\begin{eqnarray}
\bR_{\vf}&=&\bR_{\bx}-\bR_\mathrm{s} \nonumber \\
&=& \bR_{\bs}^{1/2}
(\bR_{\bs}^{-1/2} \bR_{\bx} \bR_{\bs}^{-1/2}-\bI)
\bR_{\bs}^{1/2} \nonumber \\
&=&\bR_{\bs}^{1/2}
(\bU_{\vf}(\bD_{\vf}-\bI_{\{\text{m} \}})\bU^T_{\vf})
\bR_{\bs}^{1/2} \nonumber \\
&\simeq & \bR_{\bs}^{1/2}
(\bU_{\vf}\bD_{\vf}\bU^T_{\vf})
\bR_{\bs}^{1/2}, \label{eq:Rf}
\end{eqnarray}
where the last approximate equality is because of $\lambda_{i}\gg 1$. With this approximation, we can obtain the estimated $\HI$ 21-cm emission covariance matrix
\begin{eqnarray}
\bR_{\hat{\bs}}&=& \bR_{\bx}-\bR_{\vf} \nonumber \\
&=&
\bR_{\bs}^{1/2}(\bU_{\bs}\bU_{\bs}^T)\bR_{\bs}^{1/2}\nonumber \\
&=&(\bR_{\bs}^{1/2}\bU_{\bs})(\bR_{\bs}^{1/2}\bU_{\bs})^T\,,
\label{eq:R_hat_s_est}
\end{eqnarray}
where we substituted Eqs.~(\ref{eq:eigen}) and (\ref{eq:Rf}) in the second equality. 

Using Eq.\,\eqref{eq:R_s_t^prime} and choosing $\boldsymbol{V}=\bR_{\vt^\prime}^{-1/2}$ (i.e. $\boldsymbol{V}\boldsymbol{R}_{\vt^\prime}\boldsymbol{V}^{\rm T}=\bI$), we have $\bR_{\hat{\bs}}=\bS\bS^T$, then we obtain a proper estimate of the mixing matrix
\begin{equation}
\bS=\bR_{\bs}^{1/2} \bU_{\bs}\,. \label{eq:Smat}
\end{equation}
So once we know the covariance matrix of the 21-cm signal, we can use Eq.~(\ref{eq:Smat}) to obtain the mixing matrix. An essential precondition is to figure out the degrees of freedom of the foregrounds, i.e. $m$-values, which we will elaborate in greater details in Sec.~\ref{subsubsec:dofm}.

In summary, the key step of GNILC (eGNILC) is to find the eigenvectors and eigenvalues in Eq.~\eqref{eq:eigen} to determine the mixing matrix. In Eq.~(\ref{eq:Smat}), we have only used the spatial correlations between maps to calculate the covariance matrix. But if the accuracy of component separation can be improved, the estimated 21-cm signal will be more precise. The DCT method provides such a possibility, because it helps select a better basis for the GNILC by using the frequency-smoothness of the foregrounds. We will examine the improvement in Sec.~\ref{sec:demon}.

\subsubsection{Degrees of freedom of the foregrounds}
\label{subsubsec:dofm}
Though we have assumed those eigenvalues of $\bR_{\bs}^{-1/2}\bR_{\bx}\bR_{\bs}^{-1/2}$
that correspond to the 21-cm signal are unity, practically the critical eigenvalue remains unknown to us due to imperfect estimation of the covariance matrices. Given the data $\boldsymbol{x}(p)$, we can calculate the likelihood of the mode illustrated in Sec.~\ref{subsubsec:mixing_matrix}.
To select the best value of $m$ and prevent overfitting, \citet{Olivari:2015tka} suggested the Akaike Information Criterion (AIC)
\begin{equation}
\text{AIC}(m)= 2 m -2\log(\mathcal{L}_{\mathrm{max}}(m)) \,,
\end{equation}
where $\mathcal{L}_{\mathrm{max}}(m)$ is the maximum likelihood with regard to the model parameter $m$ for the given data,
\begin{equation}
-2\log(\mathcal{L}_{\mathrm{max}}(m))= \sum_{i=1}^{n_\text{ch}-m}\left[\mu_i-\ln{\mu_i}-1\right] \, ,
\end{equation}
where we neglect the constant in the logarithmic likelihood, $\mu_i$ is the eigenvalues of $\bR_{\bs}^{-1/2}\bR_{\bx}\bR_{\bs}^{-1/2}$, and $2m$ is the penalty term that  prevents overfitting.  For details of the derivation, please refer to Appendix B of~\citet{Olivari:2015tka}.

The above AIC only explicitly uses the likelihood of $\boldsymbol{x}(p)$, but in fact, the 21-cm prior maps can also be used when calculating the likelihood because they are the ``prior knowledge'' of reconstructed signal $\hat{\boldsymbol{s}}$. Here we do not distinguish the difference between the real signal and the prior because we assume their covariance matrices are very close. We will see from the following derivation that this additional likelihood (except removable constants) is independent of the prior maps under certain assumptions.

Like the ILC bias estimation in~\citet{Delabrouille:2008qd}, we will take a similar procedure to derive the GNILC error for the 21-cm reconstruction. The reconstructed 21-cm signal $\hat{\boldsymbol{s}}$ at each pixel is
\begin{equation}
\hat{\boldsymbol{s}} = \boldsymbol{s} + \boldsymbol{\delta} \,,
\label{eq:recon_error}
\end{equation}
where $\boldsymbol{\delta}$ is the reconstruction error, and $\boldsymbol{s}$ is the underlying true signal. Thus the covariance of $\hat{\boldsymbol{s}}$ is
\begin{equation}
\hat{\bR}_{\bs}=\bR_{\bs} + 2 \boldsymbol{C}_{\boldsymbol{s\delta}} + \bR_{\boldsymbol{\delta}} \,,
\label{eq:R_hat_s_dof}
\end{equation}
where $\hat{\bR}_{\bs}$, $\bR_{\bs}$ and $\bR_{\boldsymbol{\delta}}$ are the covariance matrices of each term in Eq.~(\ref{eq:recon_error}). $\boldsymbol{C}_{\boldsymbol{s\delta}}$ is the covariance between the real 21-cm signal and the reconstruction error.
We should notice that $\hat{\bR}_{\bs}$ is different from that in Eq.~\eqref{eq:R_hat_s_est}, because the latter is an intermediate estimation to calculate mixing matrix.

We assume to reach a small GNILC error $\boldsymbol{\delta}$ such that the $\bR_{\boldsymbol{\delta}}$ term can be neglected in Eq.~\eqref{eq:R_hat_s_dof}.
This assumption is reasonable because we will achieve a low reconstruction bias after the foreground removal. The $\boldsymbol{C}_{\boldsymbol{s\delta}}$, which couples the error to the signal and leads to signal loss, is of interest to us.
For the sake of neatness, the signals are assumed/transformed to a normal distribution with mean zero. Because the covariance remains unchanged, we will not distinguish the differences in mean values, so we have $\boldsymbol{C}_{\boldsymbol{ab}} = E(\boldsymbol{a} \boldsymbol{b}^\mathrm{T})$, where $E$ denotes the ensemble average.
From Eq.~\eqref{eq:recon_error}, we can calculate the error as
\begin{eqnarray}
\boldsymbol{\delta}&=&\hat{\boldsymbol{s}} - \boldsymbol{s} =\bW (\bs + \boldsymbol{f}) - \bs = \bW \boldsymbol{f} \nonumber\\
 &=& \bS(\bS^\mathrm{T}\hat{\bR}^{-1}_{\boldsymbol{x}}\bS)^{-1}\bS^\mathrm{T}\hat{\bR}^{-1}_{\boldsymbol{x}}\boldsymbol{f} , \label{eq:delta_calculation}
\end{eqnarray}
where we have utilized Eqs.~\eqref{eq:x_s_f} and ~\eqref{eq:extract_s} in the second equality, and Eq.~~\eqref{eq:mixing} in the fourth equality. The third equality is due to the unbiasedness of the operator $\bW$ (i.e. $\bW \bs = \bs$). $\hat{\bR}_{\bx}$ is the estimation of $\bR_{\bx}$ by calculating it in a domain $\mathcal{D}$ around a pixel $p$, which is related to the true covariance matrix $\bR_{\bx}$ via $\hat{\bR}_{\bx}= \bR_{\bx} + \bDelta_{\bx}$. Thus, $\bDelta_{\bx}$ accounts for the error of the $\bR_{\bx}$ matrix. $\hat{\bR}^{-1}_{\boldsymbol{x}}$ can be further expanded to the first order as
\begin{eqnarray}
\hat{\bR}^{-1}_{\boldsymbol{x}} \simeq \bR^{-1}_{\boldsymbol{x}} -\bR^{-1}_{\boldsymbol{x}} \bDelta_{\bx} \bR^{-1}_{\boldsymbol{x}} . \label{eq:Rxhat_inverse}
\end{eqnarray}

Substituting Eq.~(\ref{eq:Rxhat_inverse}) into Eq.~(\ref{eq:delta_calculation}), we have
\begin{eqnarray}
\boldsymbol{\delta} &=& \bS\left[\left(\bS^\mathrm{T}\bR^{-1}_{\boldsymbol{x}}\bS \right)^{-1} +  \left(\bS^\mathrm{T}\bR^{-1}_{\boldsymbol{x}}\bS\right)^{-1} \right. \nonumber \\ 
&\times & \left.  \left(\bS^\mathrm{T}\bR^{-1}_{\boldsymbol{x}}\bDelta_{\bx} \bR^{-1}_{\boldsymbol{x}}\bS\right) \left(\bS^\mathrm{T}\bR^{-1}_{\boldsymbol{x}}\bS\right)^{-1}
\right] \nonumber \\
&\times &\bS^\mathrm{T} \left(\bR^{-1}_{\boldsymbol{x}} -\bR^{-1}_{\boldsymbol{x}} \bDelta_{\bx} \bR^{-1}_{\boldsymbol{x}}\right) \! \boldsymbol{f} \nonumber \\
&=& \bS \left(\bS^\mathrm{T}\bR^{-1}_{\boldsymbol{x}}\bS\right)^{-1} \bS^\mathrm{T} \bR^{-1}_{\boldsymbol{x}} \boldsymbol{f} \label{eq:d_p1}\\
& - & \bS \left(\bS^\mathrm{T}\bR^{-1}_{\boldsymbol{x}}\bS\right)^{-1} \bS^\mathrm{T} \bR^{-1}_{\boldsymbol{x}} \bDelta_{\bx} \bR^{-1}_{\boldsymbol{x}}\! \boldsymbol{f} \label{eq:d_p2}\\
& +& \bS \left(\bS^\mathrm{T}\bR^{-1}_{\boldsymbol{x}}\bS\right)^{-1} \left(\bS^\mathrm{T}\bR^{-1}_{\boldsymbol{x}}\bDelta_{\bx} \bR^{-1}_{\boldsymbol{x}}\bS\right) \nonumber \\
& \times &\left(\bS^\mathrm{T}\bR^{-1}_{\boldsymbol{x}}\bS\right)^{-1} \bS^\mathrm{T} \bR^{-1}_{\boldsymbol{x}}\! \boldsymbol{f} \label{eq:d_p3}\\
& +& \boldsymbol{\mathcal{O}}(\Delta_\mathrm{x}^2) \boldsymbol{f} \label{eq:d_p4},
\end{eqnarray}
where Eq.~\eqref{eq:d_p4} is negligible because it is the second order of $\bDelta_{\bx}$. The correlation $\boldsymbol{C}_{\boldsymbol{s\delta}} = \mathrm{E}(\boldsymbol{\delta}\boldsymbol{s}^T)$ comprises correlations from Eqs.\,\eqref{eq:d_p1} to \eqref{eq:d_p3}, and Eq.~\eqref{eq:d_p1} will vanish because $\bs$ and $\boldsymbol{f}$ are uncorrelated.

The covariance matrix $\hat{\tilde{\bR}}$ of $\bs$ and $\vf$ is calculated in a domain $\mathcal{D}$ around a pixel $p$ in the following equation. We first calculate the error $\bDelta_{\bx}$ at the pixel $p$ 
\begin{eqnarray}
\bDelta_{\bx} (p) &=& \hat{\bR}_{\bx} - \bR_{\bx} \nonumber \\
&=& \frac{1}{N_p}\sum_q(\bs_q + \boldsymbol{f}_q)(\bs_q + \boldsymbol{f}_q)^\mathrm{T} - (\bR_{\bs} + \bR_{\vf}) \nonumber \\
&=& \tilde{\bR}_{\bs} - \bR_{\bs} + \tilde{\bR}_{\vf} - \bR_{\vf} \nonumber \\
&& + \frac{1}{N_p}\sum_q(\bs_q \boldsymbol{f}_q^\mathrm{T} + \boldsymbol{f}_q \bs_q^\mathrm{T}) , \label{eq:Delta_xp}
\end{eqnarray}
where $N_p$ is the effective number of independent pixels in domain $\mathcal{D}$, and $q$ sums over all pixels. $\tilde{\bR}_{\bs}$ is defined as $\tilde{\bR}_{\bs}=(1/N_{p})\sum \bs_q \bs^{\rm T}_q$ and the same for $\tilde{\bR}_{\vf}$.

Through a lengthy calculation of the actual signal covariance matrix (Appendix~\ref{app:AIC}), we obtain the covariance of $\hat{\boldsymbol{s}}$ as
\begin{equation}
\hat{\bR}_{\bs}= \left(1 -2 \frac{m}{N_p}\right) \bR_{\bs}\,,
\end{equation}
which is smaller than the signal covariance $\bR_{\bs}$ by a factor of $2 {m}/{N_p}$.

We can finally reach the  AIC value of the likelihood with $m-$modes of foreground and $N_{\rm p}$ pixels as:
\begin{eqnarray}
\mathrm{AIC}(m, N_\mathrm{p})&=& 2 m + n_\mathrm{ch}\left[\frac{1}{1-2m/N_{\rm p}}+ \ln\left(1-2\frac{m}{N_\mathrm{p}} \right)\right] \nonumber \\
&+& \sum_{i=1}^{n_\text{ch}-m}\left[\mu_i-\ln{\mu_i}-1\right] \,,
\label{eq:AIC_mod}
\end{eqnarray}
where the additional term is scaled by the channel number $n_\mathrm{ch}$. It is not an explicit function of the 21-cm prior but the number of channels and the number of independent pixels are involved.
The effect of this additional term is not noticeable if the term $m/N_{\rm p}$ is small (most cases), but the modified AIC might be distinguished for large $n_\mathrm{ch}$.

\subsection{RPCA blind foreground removing}
\label{subsec:RPCA}
In Sec.\,\ref{subsubsec:mixing_matrix}, we assume the covariance matrix of 21-cm signal is computed through known maps, which means that the GNILC depends on the prior power spectrum of $\HI$.
It is verified that the GNILC method is not critically sensitive to the absence of detailed features in the prior HI power spectrum but depends on the overall amplitude of the $\HI$ power spectrum~\citep{Olivari:2015tka}.
Under particular circumstances where the robust principal component analysis (RPCA) is efficient, the mixing matrix $S$ can be derived from the cross-correlation matrix of data directly without the input of simulated $\HI$ 21-cm maps.
\cite{Zuo:2018gzm} implemented RPCA to extract the $\HI$ 21-cm signal, based on the low-rank property of the foreground cross-correlation matrix and the sparsity of the $\HI$ 21-cm cross-correlation matrix.
Here, we can apply the RPCA method within the GNILC framework and establish a blind method that retains the merits of the GNILC method.
Let matrix $\bR_\mathrm{L}$ and $\bR_\mathrm{S}$ be the low-rank and sparse components of the data covariance matrix $\bR_{\bx}$ found by the RPCA algorithm, we obtain the estimation of 21-cm cross-correlation matrix $\bR_{\hat{\bs}}=\bR_\mathrm{S}$.
With this replacement, we take the standard steps of GNILC to extract the $\HI$ 21-cm signal.
We test its performance with simulated data, and show the demonstration in Sec.\,\ref{sec:demon}.

The noise effect is not tested in the RPCA-embedded GNILC approach. When extracting the \HI signals from the 21-cm IM data, the larger the thermal noise amplitude is, the less accurate the reconstruction of the \HI power spectrum on small angular scales is~\citep{Olivari:2015tka}. A solution is also suggested in the case that we have a good estimate of the thermal noise power spectrum. The \HI + noise covariance matrix is used as the prior instead of the \HI covariance matrix, so that we recover the \HI signal plus thermal noise. We may correct for the thermal noise by using the estimate of the instrumental noise power spectrum. This correction can also be applied to the RPCA-embedded GNILC approach where the prior covariance is replaced with the data covariance splitting. The sparse matrix comprises the \HI covariance and the thermal covariance (diagonal for simple models), which is roughly equivalent to the \HI plus noise prior. So we expect the performance of the RPCA-embedded GNILC approach is close to the GNILC with \HI signal plus thermal noise prior.

\subsection{Summary of the eGNILC}
\label{subsec:refine}
This section summarizes the eGNILC method we developed in this paper. We will customize the GNILC to the 21-cm IM and present the general procedure of the 21-cm foreground removal. Unlike applying the internal linear combination to the CMB data set, we take additional steps to determine the mixing matrix because the 21-cm signal varies with frequency. 
\begin{itemize}
    \item The 3D 21-cm tomography data to be cleaned is projected onto a 2D sphere, and the data array at each pixel is arranged as a vector indexed by frequency. A universal spherical mask is applied to cut off extreme foreground contamination.
    The data vector is therefore denoted by $\boldsymbol{x}(p)$, where $p$ is the spatial index.
    \item We are free to decompose the vector with the ``Fourier'' transform, e.g. the data is represented by the coefficients of DCT modes in Fig.\,\ref{fig:sim_masked_maps}. Then the real map at each ``frequency'' is transformed to a needlet space where localization in both the spatial domain and the spherical harmonics domain is reserved. As a consequence, the indices $i$ and $p$ of elementary data $x_i(p)$ do not necessarily represent the frequency and pixel in real space. 
    \item The ILC algorithm is implemented in the general needlet domain. \\
    \\
    It is crucial that we figure out the imaginary mixing matrix of 21-cm signals. For 21-cm experiments, we do not know the {\it dof} of the 21-cm signal nor foregrounds, but expect that the foregrounds are smooth along frequency. With the help of simulated 21-cm signals, this goal is achievable as is done in Sec.\,\ref{subsubsec:mixing_matrix}, where simulated 21-cm signals are preprocessed in the same way as $x_i(p)$.
    However, finding the parameter $m$ ({\it dof} of foregrounds) is an essential condition to estimate the mixing matrix.
    The $m$ is determined by minimizing the Akaike Information Criterion. We derive the GNILC bias and take the size of the averaging domain into account. So the AIC is modified as in Eq.~(\ref{eq:AIC_mod}).  
    With no prior knowledge, we can also estimate the mixing matrix from data $\boldsymbol{x}(p)$ with the RPCA to get a blind method, as elaborated in Sec.\,\ref{subsec:RPCA}.
    \item After extracting signals with ILC on each needlet base, we reverse the decomposition and sum over all modes to recover the 21-cm signal.
\end{itemize}

\section{Demonstration}
\label{sec:demon}

\begin{figure*}[htbp!]
  \centering
  \includegraphics[width=0.95\textwidth,keepaspectratio]{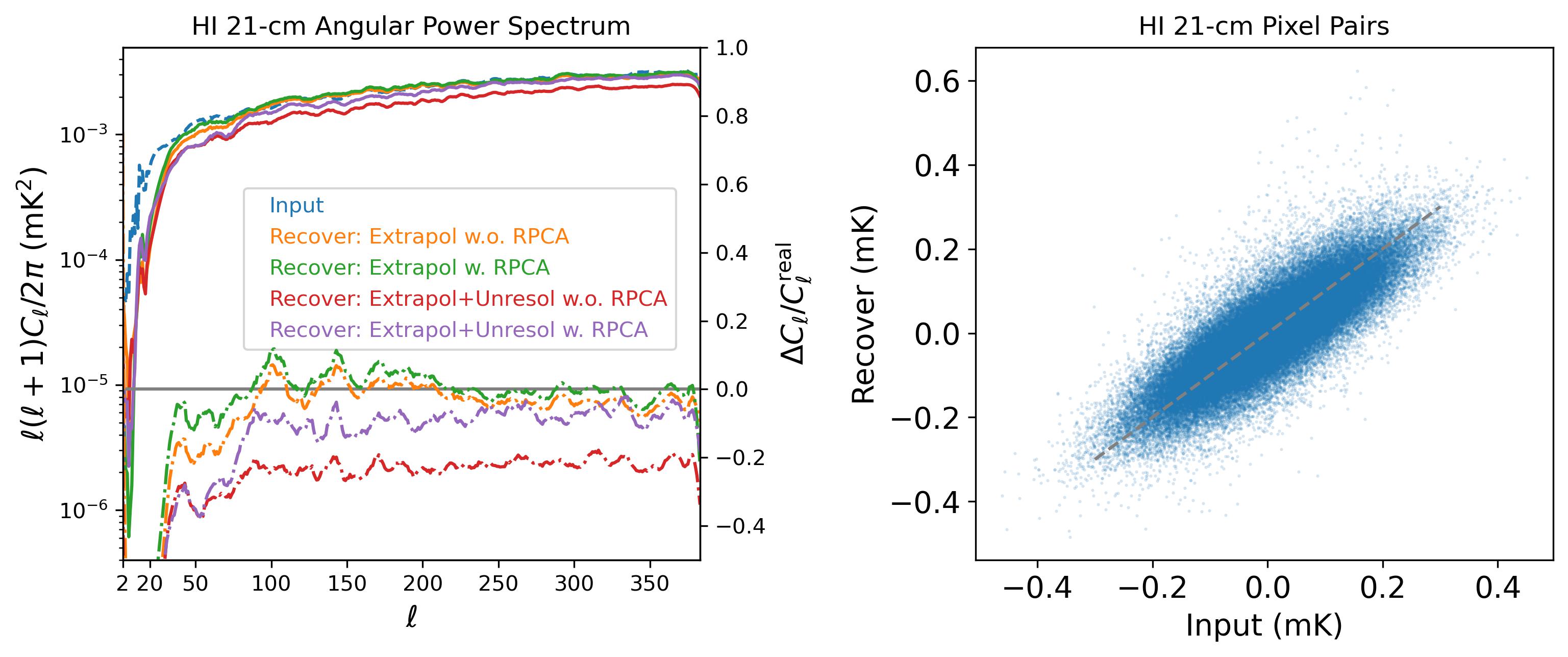}
  \caption{Demonstrations of the RPCA and foreground effects on the GNILC foreground removal method with $20$ channels. The left panel shows the 21-cm angular power spectrum at $1252.5\,\mathrm{MHz}$, where the blue dashed line is the input power spectrum, the orange (green) solid line is the power spectrum recovered from ``Extrapolation'' power-law foregrounds without (with) RPCA. The red (purple) solid line is recovered from ``Extrapolation'' and ``Unresolved'' foregrounds combined without (with) RPCA. The dotted-dashed lines in the lower region with the same colors are the corresponding fractional differences of power spectra ($\Delta C_{\ell}/C^{\rm real}_{\ell}$) with respect to the input. The right panel shows a pixel-by-pixel comparison ($\sim 100,000$ pixels) between the input map and the recovered map with the ``Extrapolation'' foreground model but without the RPCA method (its power spectrum is shown in the orange line on left panel).}
  \label{fig:demon}
\end{figure*}

\begin{figure*}[htbp!]
  \centering
  \includegraphics[width=0.95\textwidth,keepaspectratio]{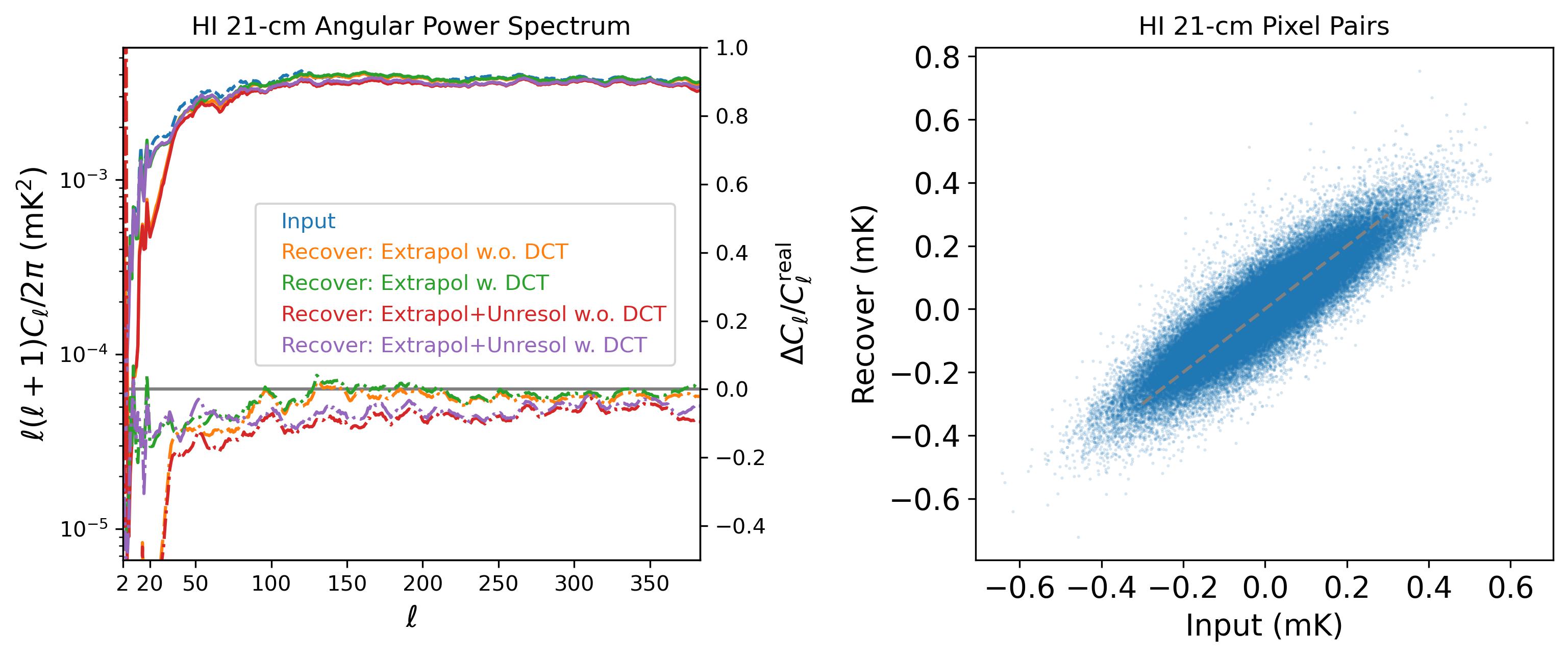}
  \caption{Same as Fig.~\ref{fig:demon} but for DCT method and $60$ channels in the same frequency range.}
  \label{fig:demon_dct}
\end{figure*}

The basic assumption of the GNILC method is that the foregrounds are highly correlated over frequencies so that the majority of the {\it dof} of $\bR_{\bx}$ encodes the $\HI$ signal.
Figure~\ref{fig:demon} shows the recovered $\HI$ power spectrum and the effects of increasing the complexity of foreground emissions.
We also demonstrated the performance of the RPCA-embedded GNILC method.
In this section, we ignored all the instrumental effects (e.g., thermal noise and beams).

\begin{figure}
  \centering
  \includegraphics[width=0.48\textwidth,height=0.3\textwidth]{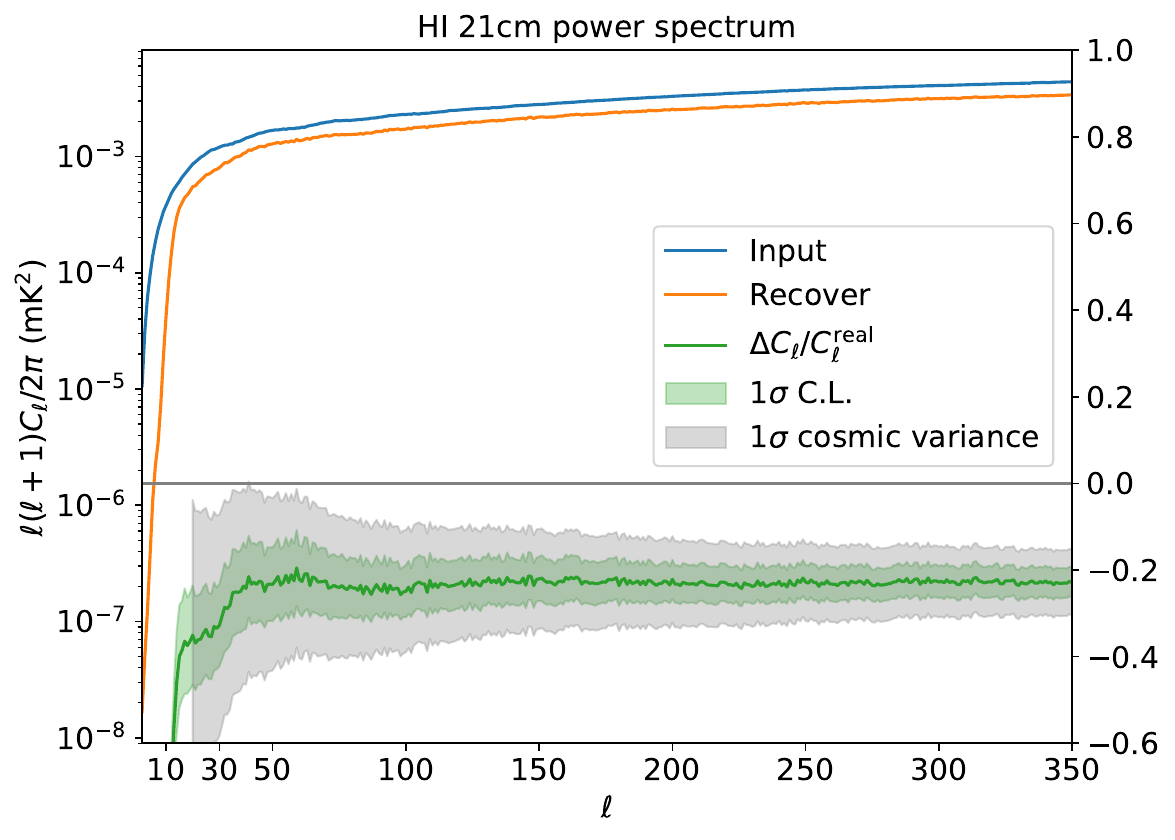}
  \caption{Fluctuations of the recovered power spectrum by the plain GNILC in $1000$ realizations of $\HI$ 21-cm intensity maps, with the same foreground model and mask as in Fig.\,\ref{fig:sim_masked_maps}. The simulation runs with $20$ frequency channels and contains power-law and unresolved foregrounds. The cosmic variance (gray band) is calculated as $\Delta C_{\ell}/C_{\ell}=\sqrt{2/[f_\mathrm{sky}(2\ell +1)]}$, where $f_\mathrm{sky}=0.5$ is the sky coverage factor.} 
  \label{fig:stochastic}
\end{figure}

We simulate foregrounds and $\HI$ signals in the frequency range $960\,\mathrm{MHz}$-$1260\,\mathrm{MHz}$ with $20$ channel bins evenly spaced.
We switch the foreground between ``Extrapolation'' and ``Extrapolation + Unresolved'', the method between ``RPCA'' and ``Simulation'' to examine the foreground cleaning effect. Figure~\ref{fig:demon} shows the recovered 21-cm auto-power spectra centered at $1252.5\,\mathrm{MHz}$ in four cases.
``Extrapolation'' denotes the well-modeled sources in Sec.~\ref{subsec:extrapolation_sky}, and ``Unresolved'' denotes the unresolved random sources shown in Sec.~\ref{subsec:unresolved_sky}.
The GNILC method and the RPCA-embedded GNILC method are utilized in the foreground removal, which are denoted by ``w.o. RPCA'' and ``w. RPCA'' separately according to the manner in which the covariance matrix is computed. Because almost all 21-cm IM experiments observe the partial sky, in the following comparison, we just compare the pseudo-power spectrum (L.H.S. of Eq.~(\ref{eq:pcl})) directly, without inverting the $M$-matrix (Eq.~(\ref{eq:Mell})) in Eq.~(\ref{eq:pcl}) to recover the full-sky power spectrum. In a word, we will suppress the tilde in $\tilde{C}_\ell$ in the following presentation.

Figure~\ref{fig:demon}'s left panel shows the 21-cm angular power spectrum at $1252.5\,\mathrm{MHz}$, where the blue dashed line is the input power spectrum, and the orange and green solid lines are the power spectrum recovered from ``Extrapolation'' power-law foregrounds without and with RPCA respectively. The red solid line is the recovered power spectrum from ``Extrapolation'' and ``Unresolved'' foregrounds without combining with RPCA, and the purple solid line is that combined with RPCA. In the lower region of the left panel, the dotted-dashed lines in the same color are the corresponding fractional differences of power spectra with respect to the input power spectrum. For clarity of presentation, all power spectra curves are smoothed properly to remove the small wiggles in $\ell$-space.
The right panel makes a pixel-by-pixel comparison between the recovered $\HI$ map and the input map, in which the recovered map is the case of ``Extrapolation'' power-law foregrounds without RPCA. The gray dashed line is the $y=x$ diagonal line, which indicates the unbiased estimate. One can see that the recovered and input values are highly aligned alongside of the $y=x$ line, which demonstrates the level of unbiasedness in the real space (the left panel is the Fourier space).

In the lower region of the left panel of Fig.~\ref{fig:demon}, one can see that the level of $\HI$ power spectrum underestimation is almost zero if the contaminant is just the ``Extrapolation'' foreground. But this situation deteriorates if the ``Unresolved'' foregrounds are included, in which case, better frequency resolution would be necessary to extract the signal if the complexity of foregrounds increases.
The RPCA-embedded blind GNILC method has a fairly good performance as long as the frequency resolution is good enough for foreground removal.
We notice that the underestimation of the ``RPCA-embedded'' recovered 21-cm auto-power spectrum is smaller. This does not mean that the RPCA-estimated mixing matrix is better than that with the simulated $\HI$ signal, because the unresolved foreground power contributes to the sparse matrix in the robust principal component analysis that will be used as the input covariance of the 21-cm signals.
Besides, the RPCA takes a long time to split the low-rank and sparse components, especially when the matrix size gets larger.
Moreover, the domain $\mathcal{D}$ may not be the best in this case.
From the left panel of Fig.~\ref{fig:demon}, one can see that in the range of $\ell \leq 20$, the GNILC method is not suitable to clean the foreground because of the inadequate sampling on large scales.

The HI signals and foregrounds can be decomposed via the DCT as shown in Fig.\,\ref{fig:sim_masked_maps}.
Because the frequency transform is independent of the spatial needlet decomposition, we can do the DCT of frequency before the GNILC to generalize the cleaning algorithm further.
Figure~\ref{fig:demon_dct} shows the effect of whether the DCT is applied beforehand.
The power spectra are presented in the same way as in Fig.~\ref{fig:demon}, but we change the number of frequency channels from $20$ to $60$.
Foregrounds are allowed to vary between ``Extrapolation'' and ``Extrapolation + Unresolved''; the DCT may be switched between ``ON'' and ``OFF''.
The DCT helps to improve the 21-cm signal recovery at low $\ell$s where the GNILC itself lacks accuracy. This is clear for both the “Extrapolation" and the “Extrapolation + Unresolved” foregrounds, i.e., the comparison of the blue to orange and the purple to red curves.
From Fig.\,\ref{fig:demon} to Fig.\,\ref{fig:demon_dct}, we see that with $60$ channels, the recovered power spectrum and pixel-by pixel comparisons are improved.

We also studied the stability of the GNILC method. Because the blind method is comparable to the plain GNILC method regarding accuracy, we only estimated the fluctuations of the plain GNILC method to save computing time.
The sky map is generated with fixed foreground maps, and the 1000 $\HI$ map samples are generated from the same power spectrum.
Then the $\HI$ signal is extracted with the GNILC method from each sample.
In Fig.\,\ref{fig:stochastic}, the blue curve is the theoretical angular power spectrum of $\HI$ (input), the yellow curve is the mean value of the recovered angular power spectrum of $\HI$, and the green curve and band are the difference and standard deviation.
We calculated the cosmic variance $\Delta C_\ell/C_\ell=\sqrt{2/[f_\mathrm{sky}(2\ell +1)]}$ (gray shadow, $f_\mathrm{sky}=0.5$ is the sky coverage factor) and put the center value on the green curve for a vision comparison.
The uncertainty caused by the GNILC method is insignificant because the fluctuation of the power spectrum does not override the cosmic variance.

\section{Application to single-dish experiments}
\label{sec:gnilc_impl}
\begin{figure*}[htbp!]
  \centering
  \includegraphics[width=\textwidth,keepaspectratio]{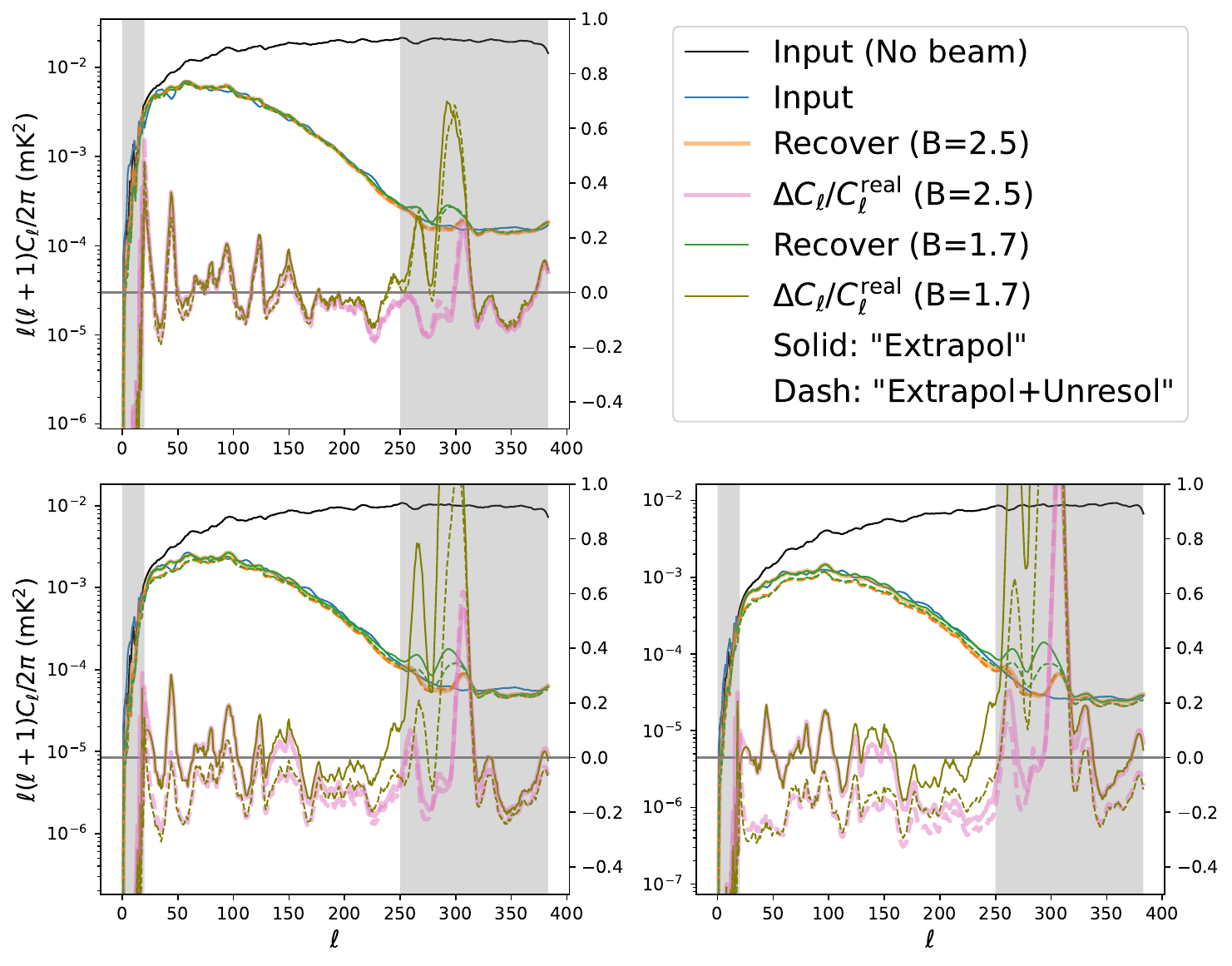}
  \caption{The recovered HI 21-cm angular power spectra of the simulated SKA-MID map with the Airy beam. The black thin line is the power spectrum of the orginal map (without beam), and the blue one is to convolve the black line with the beam (i.e. input spectrum). The orange thick curves and the green thin curves show the recovered power spectra with the spectral window function parameter $B=2.5$ and $B=1.7$, respectively. The pink thick curves and the olive thin curves show their relative differences between the input and recovered power spectrum which values can be read from the right boundary line. The solid and dashed curves represent the ``Extrapolation'' and ``Extrapolation + Unresolved'' foregrounds, respectively. The upper-left panel and lower-right panel present the auto-power spectrum at frequency $1069.9\,\mathrm{MHz}$ and $1069.7\,\mathrm{MHz}$ respectively, and the lower-left is the cross-power spectrum at these two frequencies. The left and right shadows ($\ell\lesssim 20$ \& $\ell \gtrsim 250 $) indicate the cutoff multipole ranges due to poor accuracy.}
  \label{fig:ska_mid_aps}
\end{figure*}

\begin{figure*}[htbp!]
  \centering
  \includegraphics[width=\textwidth,keepaspectratio]{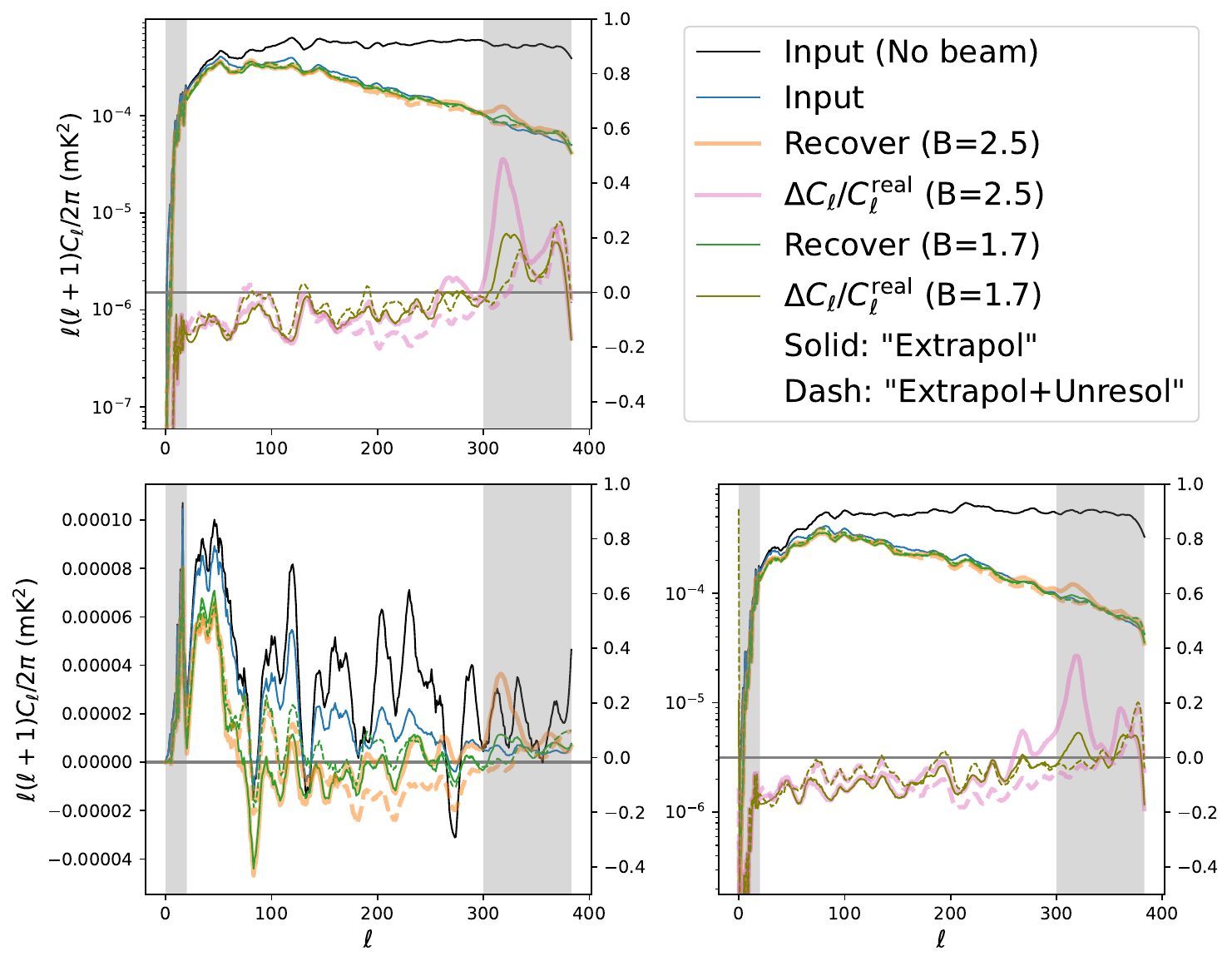}
  \caption{Same as Fig.~\ref{fig:ska_mid_aps} but for BINGO experiment. The upper-left panel and lower-right panel present the auto-power spectrum at frequency $1257.5\mathrm{MHz}$ and $1252.5\mathrm{MHz}$ respectively, and the lower-left is the cross-power spectrum (fractional differences omitted).}
  \label{fig:bingo_aps}
\end{figure*}

We implement the eGNILC foreground removal framework in the simulated data for two SD experiments, the SKA-MID and BINGO, both of which probe the low-$z$ 21-cm power spectrum. The DCT transform and the modified AIC is used, and the RPCA is switched off to save computation time. We limit the sky coverage in accordance with the experimental designs and take into account the beam effect, with experimental specifications shown in Table~\ref{tab:instru_param}. The Airy beam is truncated at $6$ times the radius of the first zero, and the radius is set by best-fitting the parameter $\theta_{\text{FWHM}}$. With this truncation, the last sidelobe is $40\,\mathrm{dB}$ below the maximum, but one should notice that the real primary beam could exhibit much more complex patterns, such as those investigated in \citet{2021MNRAS.502.2970A} for the MeerKAT (SKA-MID precursor) L-band beams. The Airy power pattern is~\citep{Condon2016} 
\begin{eqnarray}
    P(u) = \left[\frac{2J_1(\pi u D/\lambda)}{\pi u D/\lambda} \right]^2 \,,
\end{eqnarray}
where $J_1$ is the first-order Bessel function of the first kind, $u=\sin\theta$ is the sine of the zenith angle, $\lambda$ is the wavelength, and $D$ is the aperture width. Therefore, the beam size is scaled by the frequency ratio $\nu_0/\nu$ where $\nu_0$ is a reference frequency, which may cause significant errors on typical scales in the recovery. We will see abrupt peaks in the recovered angular power spectra of the SKA-MID and BINGO in Fig.\,\ref{fig:ska_mid_aps} and Fig.\,\ref{fig:bingo_aps}.
The overall system temperature consists of the receiver noise temperature, the temperature of the CMB, the galaxy temperature, etc.
As we stated in Sec.\,\ref{subsubsec:noise}, the noise effect is not the major issue of this paper, and we will only extract the 21-cm signal from the foreground contaminants.
The potential RFI contamination is also ignored.

To recover the signal with good accuracy, we should deconvolve the native beams and reconvolve a common beam to all maps~\citep{Delabrouille:2008qd}. However, in this Section, we do not deconvolve the beam from the simulated data before doing the foreground removal. Perfect beam deconvolution is not easy due to nonnegligible instrumental noise, scanning strategy and map making, and beam distortion~\citep{Burigana:2003ca,Planck:2015wtm}. In the real data, it is possible that the beam effect persists on small scales after beam deconvolution. We will present how the eGNILC works in a bad situation where  frequency-dependent Airy beams are convolved to maps and beam deconvolution is absent, although the instrument noise is not added.

\begin{figure}
  \centering
  \includegraphics[width=0.45\textwidth,keepaspectratio]{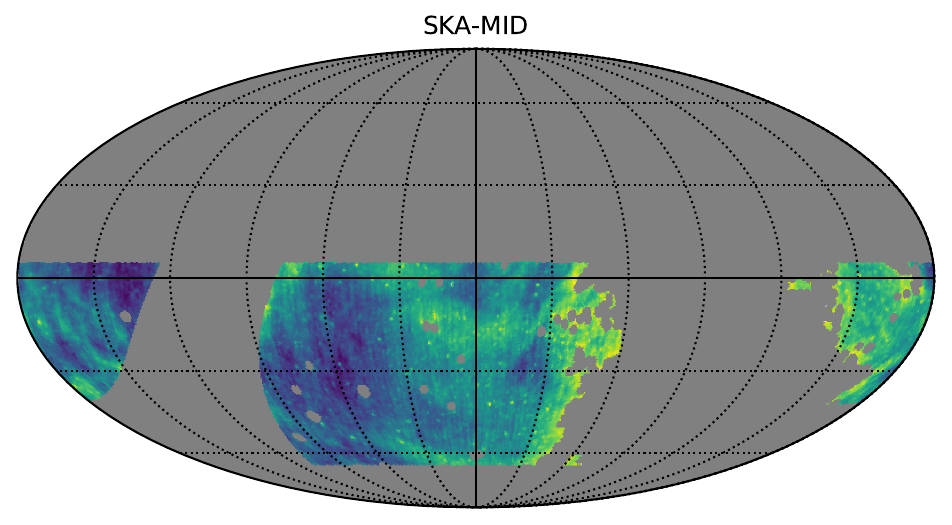}
  \includegraphics[width=0.45\textwidth,keepaspectratio]{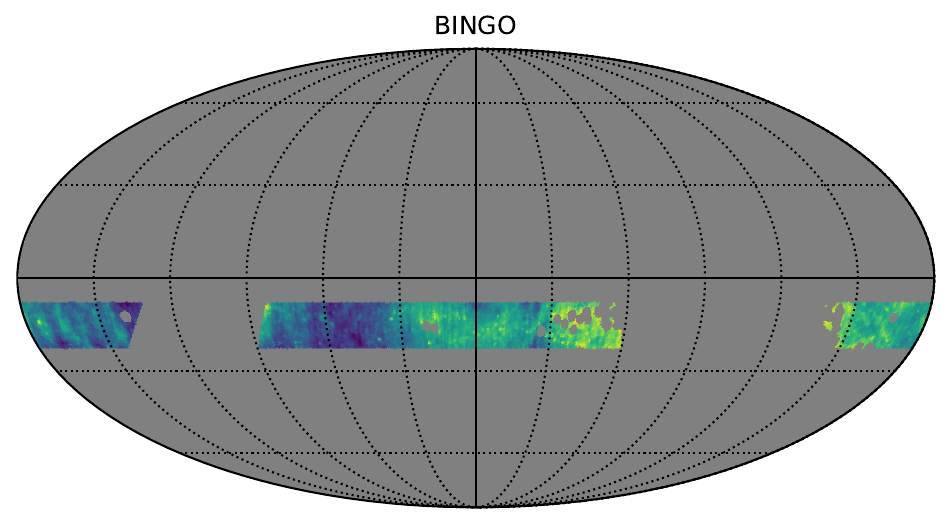}
  \caption{The sky coverage of the simulated SKA-MID and BINGO, projected in the Equatorial coordinate. The gray area is masked, which is derived from {\tt Haslam} map. The valid area is reduced from the survey area of $20,000$ to $10,000 \deg^2$ and from the survey area of $5000$ to $3000 \deg^2$ (Table~\ref{tab:instru_param}), respectively, because of the Galactic plane. See also Sec.~\ref{subsec:ska-mid}.}
  \label{fig:masks}
\end{figure}

\subsection{SKA-MID}
\label{subsec:ska-mid}
SKA-MID is a dish array consisting of a set of subarrays, which comprises the South African SKA precursor MeerKAT (launched in 2018) with 64 $13.5\mathrm{m}$-diameter dishes and the supplement of $133$ $15\mathrm{m}$-diameter SKA1 dishes~\citep{Bacon:2018dui}.
The telescopes are configured with a compact core and three logarithmically spaced spiral arms with a maximum baseline of $150$\,km, which corresponds to an angular resolution $0''.3$ at the frequency of $1.4\,\mathrm{GHz}$~\citep{Bacon:2018dui}.
However, SKA-MID in interferometric mode does not provide enough short baselines to map the scales of interest with sufficient signal-to-noise~\citep{Bull:2014rha}.
We use the single-dish mode instead, which generates the auto-correlation information from each dish\footnote{The single-dish IM is an auto-correlation of each dish's voltage, which is different from the interferometric mode where the visibility is obtained by cross-multiplying the voltage.}. The large number of dishes available with the telescopes will guarantee a large survey speed for probing the $\HI$ signal~\citep{Santos:2015gra}.

\citet{Wang:2020lkn} calibrated 10.5-hour MeerKAT L-band ($856$-$1712 \,\mathrm{MHz}$, 4096 channels) autocorrelation data, in which only the $971$-$1075\,\mathrm{MHz}$ and $1305$-$1504\,\mathrm{MHz}$ bands are free from strong RFI contamination.
In this paper, we simulate the data spanning the frequency band $970\,\mathrm{MHz}$-$1070\,\mathrm{MHz}$ with channel width $\delta \nu = 0.2\,\mathrm{MHz}$, which corresponds to the redshift range of $z=[0.33, 0.46]$. These configurations are not necessarily the same as the real telescopes, because we are using the simulated data, the actual frequency channels are not critical in studying the effects of the mask and beam in the GNILC foreground removal.
Fig.~\ref{fig:masks} illustrates the sky coverage of the SKA-MID and BINGO with highly contaminated areas (mostly the galactic disk) removed.
The sky coverage of the SKA-MID data is around $10,000 \deg^2$.
More details are listed in Table\,\ref{tab:instru_param}. 

We apply the eGNILC algorithm to the simulated SKA-MID data. The foreground consists of the ``Extrapolation'' with or without the ``Unresolved'' components. Because the frequency resolution is $\delta \nu =0.2\,$MHz (Table~\ref{tab:instru_param}), the neighboring 21-cm maps are highly correlated\footnote{One may refer to~\citet{Shaw2008} and~\citet{LiMa2017} for detailed calculations of the neighboring frequency band cross-correlation.}. This may potentially cause practical difficulties while inverting the signal covariance because of the small eigenvalues of $\bR_{\bs}$. If so, then the computational error is out of control, and consequently, the GNILC foreground removal fails in its algorithm. The solution is to separate the maps into $n$ subsets.
For the full set indexed with $\{1_1,1_2,\ldots, 1_n; 2_1,2_2,\ldots, 2_n; \cdots; i_1,i_2,\ldots,i_n; \cdots\}$, $n$ should be carefully assigned to guarantee that all maps in the subset $\{1_k,2_k,\cdots, i_k,\cdots\}$ are uncorrelated.
We set $n$ to $10$ for the $500$ frequency channels, which is adequate to break the correlation among 21-cm maps within the subsets and is large enough to obtain high recovery accuracy.
The eGNILC is applied to each subset to extract the HI signal.

Figure~\ref{fig:ska_mid_aps} presents the recovered HI 21-cm angular power spectra at $1069.9\,\mathrm{MHz}$ and $1069.7\,\mathrm{MHz}$ that belong to two different subsets. The original 21-cm maps, together with the foreground, are then convolved with the Airy-disk primary beam to mimic the measurement, and the recovered HI 21-cm maps are compared with the input maps which are the masked convolved 21-cm maps. For all maps, we multiplied the same reprocessed mask from {\tt Haslam} $408\,\mathrm{MHz}$ map~\citep{Remazeilles:2014mba}. In all three panels of Fig.~\ref{fig:ska_mid_aps}, the black thin curves are their original power spectra without the beam, and the blue ones are those convolved with the beam (input) so they are damped at large $\ell$s. The orange thick curves are the recovered power spectra with the spectral window function parameter $B=2.5$, while the green thin curves are recovered with $B=1.7$. The pink thick curves and the olive thin curves show their relative differences between the input and recovered power spectrum, whose values can be read out from the right boundary line. In addition, the 21-cm power spectra recovered from the ``Extrapolation'' and ``Extrapolation + Unresolved'' foregrounds are plotted in solid and dashed curves respectively.

From the lower left panel of Fig.~\ref{fig:ska_mid_aps}, one can see that there is a strong correlation between the 21-cm maps from two adjacent frequency channels. The difference between ``Input'' and ``Recovery'' is $10\%\sim 20\%$ (pink thick lines and olive thin lines; also see Table~\ref{tab:psdiff}), which shows the good agreement between the two signals. Therefore, the eGNILC can recover the 21-cm signal very well within $\ell=[20, 250]$. The truncated Airy beam decreases the power spectrum exponentially with a flat tail. For $\ell\gtrsim250$, the recovered 21-cm power spectra deviate from the original power spectra because the beam varies with frequency, where larger errors are observed in the recovered power spectrum.
Such a region is marked with a gray shadow and is not used in further analysis. Similarly, the left-most region $\ell\lesssim20$ is also marked with a gray shadow for its large bias due to insufficient independent samples for the eGNILC.
We change the parameter $B$ and foreground contaminants to see how the recovered 21-cm is affected. Figure~\ref{fig:ska_mid_aps} shows the power spectrum bias, and Table~\ref{tab:psdiff} gives the average absolute difference in $\ell=[20, 250]$ between the input power spectrum and the recovered power spectrum.
By changing the foregrounds from the ``Extrapol" to ``Extrapol + Unresol", we see the bias increases with the foreground complexity, though the difference is tiny in the auto-power spectra at $1069.9\mathrm{MHz}$. 
Changing $B$ from $2.5$ to $1.7$ also improves the performance of the eGNILC.

\begin{table*}[!htbp]
\renewcommand\arraystretch{1.5}
\begin{center}
\caption{Average normalized absolute difference between the input power spectrum and the recovered power spectrum of the SKA-MID and BINGO, in the multipole range $20<\ell<250$ and $30<\ell<300$ respectively.}
\begin{ruledtabular}
\begin{tabular}{ccccccccc}
\multirow{3}{*}{} &\multicolumn{4}{c}{SKA-MID} &\multicolumn{4}{c}{BINGO} \\ \cline{2-5} \cline{6-9} 
\multirow{3}{*}{}           &\multicolumn{2}{c}{Extrapol} &\multicolumn{2}{c}{Extrapol + Unresol} &\multicolumn{2}{c}{Extrapol} &\multicolumn{2}{c}{Extrapol + Unresol}\\ \cline{2-3} \cline{4-5} \cline{6-7} \cline{8-9}
\multirow{3}{*}{} &$B=2.5$ &$B=1.7$ &$B=2.5$ &$B=1.7$ &$B=2.5$ &$B=1.7$ &$B=2.5$ &$B=1.7$\\
\hline
$1069.9\mathrm{MHz}\times1069.9\mathrm{MHz}$ &$7.8\%$ &$6.9\%$ &$7.9\%$ &$7.0\%$ &$-$ &$-$ &$-$ &$-$\\
$1069.9\mathrm{MHz}\times1069.7\mathrm{MHz}$ &$9.5\%$ &$7.1\%$ &$13.7\%$ &$11.8\%$ &$-$ &$-$ &$-$ &$-$\\
$1069.7\mathrm{MHz}\times1069.7\mathrm{MHz}$ &$10.3\%$ &$6.8\%$ &$21.3\%$ &$17.6\%$ &$-$ &$-$ &$-$ &$-$\\
$1257.5\mathrm{MHz}\times1257.5\mathrm{MHz}$ &$-$ &$-$ &$-$ &$-$ &$8.9\%$ &$8.7\%$ &$10.5\%$ &$6.2\%$\\
$1257.5\mathrm{MHz}\times1252.5\mathrm{MHz}$ &$-$ &$-$ &$-$ &$-$ &$-$ &$-$ &$-$ &$-$\\
$1252.5\mathrm{MHz}\times1252.5\mathrm{MHz}$ &$-$ &$-$ &$-$ &$-$ &$8.9\%$ &$9.0\%$ &$10.0\%$ &$6.6\%$\\
\end{tabular}
\end{ruledtabular}
\label{tab:psdiff}
\end{center}
\end{table*}

\subsection{BINGO}
\label{subsec:bingo}
The BINGO telescope is a dedicated instrument designed specifially for observing the BAO signal with 21-cm intensity mapping and to provide new insight into the Universe at $z\lesssim 0.5$ with a dedicated instrument~\citep{Battye2012,Battye:2012tg,Dickinson2014,Battye2016,Wuensche:2019cdv,Yohana2019,Abdalla2022,Wuensche2022}.
The telescope consists of two $\sim 40$-meter compact mirrors with no moving part and will be built in a very low RFI site in South America.
The unique design can give an excellent polarization performance and very low sidelobe levels required for intensity mapping.
BINGO will map a $15^\circ$ strip on the southern hemisphere as the telescope drift scanning with $60$ fixed horns.

BINGO has a better spatial resolution than SKA-MID because of its larger dish.
In this paper, we simulate its performance in the frequency range of $960\,\mathrm{MHz}$-$1260\,\mathrm{MHz}$ with channel width $\delta \nu = 5 \,\mathrm{MHz}$, which corresponds to the redshift range of $z=0.13$-$0.48$. The sky coverage of the simulated BINGO data is about $3000 \deg^2$, and Figure~\ref{fig:masks} shows its valid area. We take the values of these telescope parameters as per guidelines in the previous literatures~(\citealt{Battye2012,Dickinson2014,Battye2016,Bacon:2018dui,Wuensche:2019cdv,Yohana2019,Wang:2020lkn,Abdalla2022,Wuensche2022}; see also Table~\ref{tab:instru_param}), but it just serves as a reference to present the characteristics of the two telescopes. The final telescope can have different experimental specifications than the values quoted here. Figure~\ref{fig:bingo_aps} presents the results of foreground removal and 21-cm signal recovery for BINGO in the same way as Figure~\ref{fig:ska_mid_aps}. Because BINGO's frequency resolution is $\delta \nu =5$\,MHz, the neighbouring 21-cm maps are almost uncorrelated, which shows in the lower-left panel that the cross-correlation is an order of magnitude lower than the auto-correlation (upper and right panels of Fig.~\ref{fig:bingo_aps}).
The cross-power spectrum is therefore plotted linearly and the relative difference is not presented because of the smallness of the cross powers.
Because the resolution of BINGO is higher, the cutoff range due to discordant beams at high $\ell$ shrinks to $\ell\gtrsim300$.
After $B$ is changed from $2.5$ to $1.7$, the average absolute difference and the abrupt peaks decrease, although the peaks are still distinct (see Table\,\ref{tab:psdiff} and Fig.\,\ref{fig:bingo_aps}).
Unexpectedly, the 21-cm signal recovered from the ``Extrapol + Unresol" has the smallest bias.
This indicates that the performance of the eGNILC rest on many key factors, e.g., mask, beam, foreground, spectral window function, and covariance calculation. Given an observation, we may vary the tuning parameters of the eGNILC to obtain the best performance.

\section{Conclusions}
\label{sec:conclusions}
In this paper, we developed the expanded generalized needlet internal linear combination method (eGNILC) by refining the algorithm of taking a discrete cosine transform beforehand, modifying the criterion for determining the {\it dof} of foregrounds, and embedding the Robust Principal Component Analysis (RPCA) in mixing matrix computation.
The eGNILC implements the ILC component separation method in a needlet space and is customized to 21-cm IM experiments.
The map is localized in both the spatial domain and the spherical harmonics domain. Furthermore, the data vector can be decomposed along the frequency axis with the ``Fourier'' transform, e.g., the data is represented by the coefficients of the DCT modes in this paper.
The robustness of such customization is demonstrated in Sec.\,\ref{sec:demon}.
Unlike the sole mode of CMB, the recovered 21-cm signals at different frequency channels are mixed virtually by an imaginary mixing matrix.
It is crucial to figure out this mixing matrix to extract the true signal successfully.
This can be accomplished with the help of simulated 21-cm signals, and knowing the parameter $m$ ({\it dof} of foregrounds) is an essential prerequisite. The $m$ is determined by minimizing the Akaike Information Criterion.
In the meantime, we derive the eGNILC bias which is unique to the 21-cm data set.
We find that it is not particular to a given map, but related to the choice of averaging domain that estimates the covariance matrix of data.
Such bias is taken into the consideration of the total likelihood of the AIC to account for the choice of averaging domain, though it is not significant in most cases.

The GNILC method uses the relative power of the $\HI$ signals to the observations to estimate the $\HI$ subspace (see also~\citealt{Olivari:2015tka}). Thus it is not sensitive to details of the prior $\HI$ maps.
Nevertheless, a prior for the $\HI$ power spectra is necessary.
We discussed a variant of GNILC with the prior knowledge of HI 21-cm signals substituted with the RPCA to compute the covariance of the $\HI$ signal.
Thus, we obtained a blind non-parametric method in the GNILC framework.
We tested the methods' response to the smooth foreground assumption and found that the RPCA-embedded GNILC performed fairly well in our demonstration. However, we should notice that the RPCA takes a long time to split the low-rank and sparse components, especially if the matrix size becomes larger. One should remember that the ILC bias is derived from the artificial anti-correlation between the signal and contaminants due to limited samples in the domain $\mathcal{D}$, which has no relation with the error caused by the split of low-rank and sparse components in the RPCA. Thus we find the caveat that the modified AIC should not be used together with the RPCA in the eGNILC framework.

We demonstrated that the fluctuation of the recovered power spectrum is smaller than the cosmic variance.
The eGNILC method was tested in more complex mock data that were simulated according to configurations for the SKA-MID and BINGO single-dish experiments.
The coverage areas are plotted in Fig.\,\ref{fig:masks}, and the effects of the telescope beam are also tested.
In tests for these two experiments, the Airy-disk beam was applied to the data and then deconvolved after foreground removal.
There are $500$ channels in the range $970\,\mathrm{MHz}$-$1070\,\mathrm{MHz}$ for the SKA-MID ($\delta \nu=0.2$\,MHz) while there are only $60$ channels in the range $960\,\mathrm{MHz}$-$1260\,\mathrm{MHz}$ for the BINGO ($\delta \nu=5\,$MHz).
The procedures of foreground removal for the SKA-MID and BINGO are different because the 21-cm maps are highly correlated in the former, while they are nearly uncorrelated in the latter.
Although the eGNILC takes advantage of the increasing {\it dof} of foregrounds to recover the 21-cm signals better, it could fail if the adjacent frequency channels are highly correlated because of the computational error of inverting the highly correlated signal covariance. We avoided this problem by grouping frequency channels in Sec.\,\ref{subsec:ska-mid}.

With the Galactic plane masked and no telescope beam, the eGNILC bias is negligible for simple power law foregrounds.
We find that the varying Airy beam leads to significant bias in the power spectrum at typical scales.
After cutting off the low-$\ell$ and high-$\ell$ multipoles limited by the number of observable samples and the telescope resolution, respectively, the test shows the customized GNILC method is applicable to the SKA-MID and BINGO single-dish experiments. Ignoring the system's noise, the \HI signals of the SKA-MID and BINGO are effectively recovered at the multipoles in $[20, 250]$ and $[20, 300]$ respectively. With $B=2.5$ or $B=1.7$, the eGNILC is applied to the experiments to remove the ``Extrapol" and ``Extrapol + Unresol" foregrounds. The SKA-MID exhibits $\lesssim 20\%$ power loss and BINGO exhibits $\sim 10\%$ power loss. The results also suggest that beam effects should be dealt with carefully, especially at high-$\ell$s.
\section{Acknowledgements}
\label{sec:ack}
We thank the useful comment from Prof. C. Dickinson. Y.Z.M. is supported by the National Research Foundation of South Africa under grant Nos. 150580, 159044, ERC23040389081 and CHN22111069370. This work was part of the research programme ``New Insights into Astrophysics and Cosmology with Theoretical Models Confronting Observational Data'' of the National Institute for Theoretical and Computational Sciences of South Africa. 

\appendix

\section{Needlet analysis}
\label{sec:app_needlet}

\begin{figure}
\begin{center}
\includegraphics[width=0.68\textwidth,keepaspectratio]{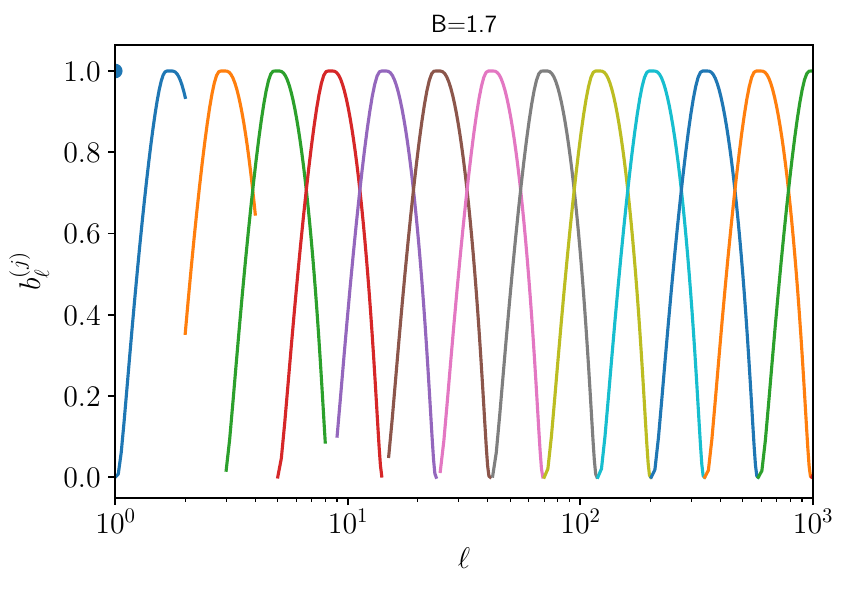}
\caption{The window function $b^{(j)}_{\ell}$ as an example of $B=1.7$.}
\label{fig:apdix_phi_example_gnilc}
\end{center}
\end{figure}

\begin{figure}
\begin{center}
  \subfloat{
  \includegraphics[width=0.42\textwidth,keepaspectratio]{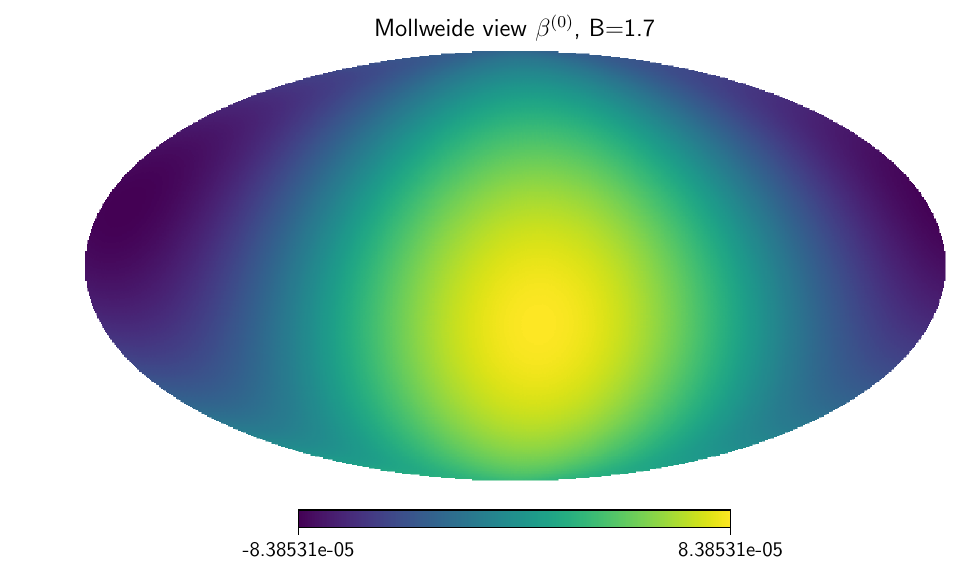}
  }
  \subfloat{
  \includegraphics[width=0.42\textwidth,keepaspectratio]{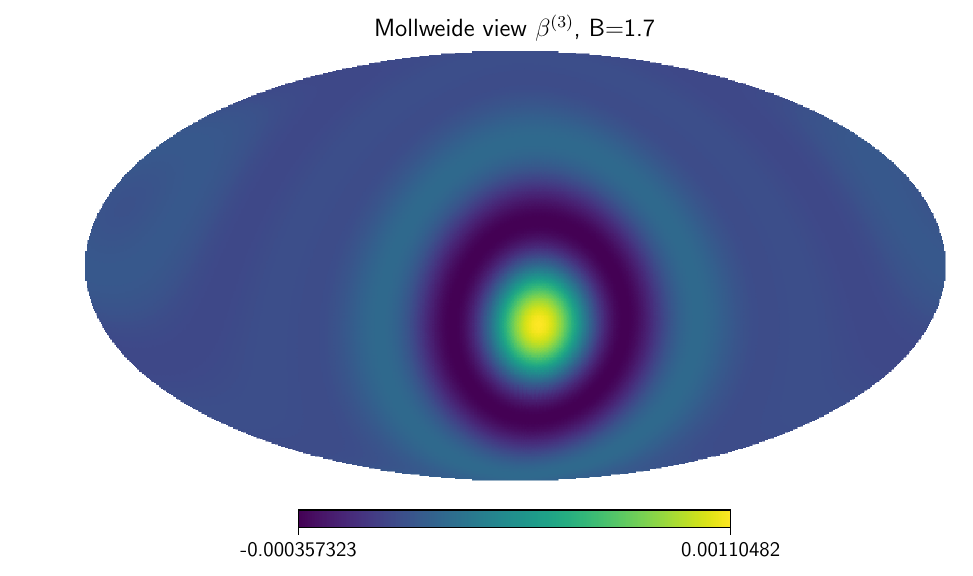}
  }
\\
  \subfloat{
  \includegraphics[width=0.42\textwidth,keepaspectratio]{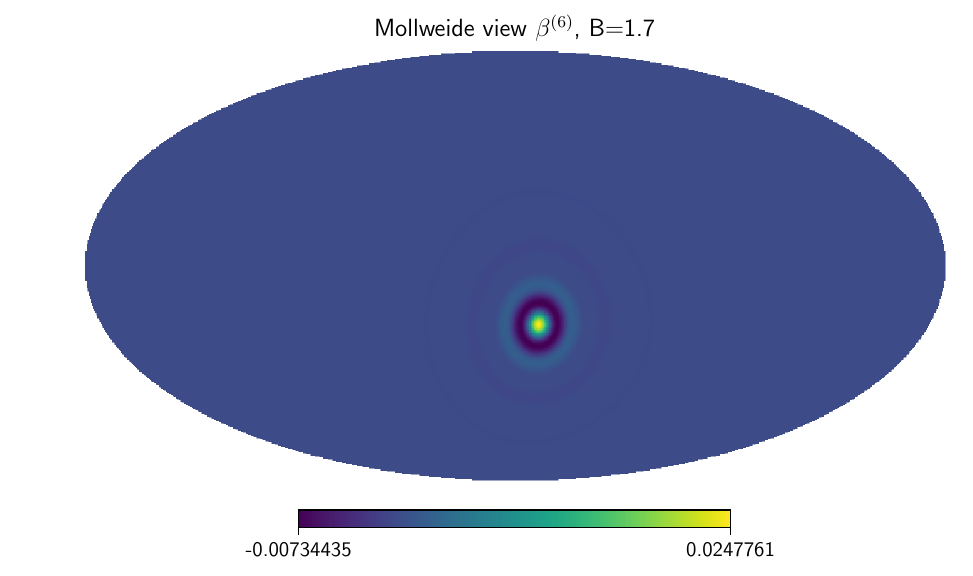}
  }
  \subfloat{
  \includegraphics[width=0.42\textwidth,keepaspectratio]{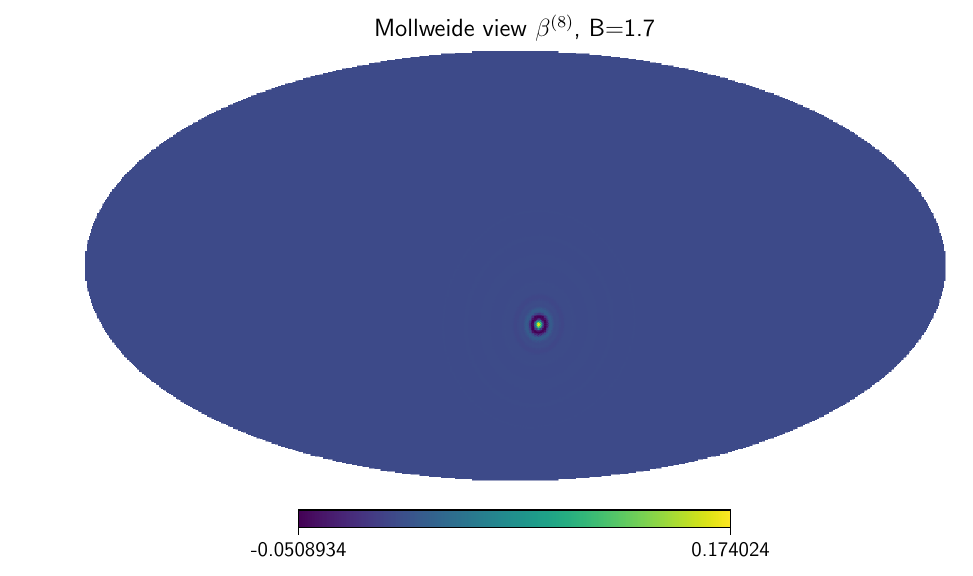}
  }
\caption{Mollweide-projected view of the needlet decomposition of a point source.}
\label{fig:apdix_point_B_adic}
\end{center}
\end{figure}

Like the Fourier transforms on the Euclid plane, the spectral analysis of data on the sphere can be conducted through spherical harmonics transform.
In many situations, we need to localize the analysis in both the spatial and spherical harmonics domains~\citep{Guilloux:2007gh}.
This requirement can be achieved through needlet analysis, which is an analogy to wavelet transform on the Euclid plane.

The usual complex spherical harmonics $(Y_{\ell m})$ ($\ell \ge 0$ and $-\ell \le m \le \ell$) on the unit sphere (denoted $\mathbb{S}$) form an orthonormal basis of the space (denoted $\mathbb{H}$) of complex-valued square-integrable functions on $\mathbb{S}$ under the Lebesgue measure ${\rm d}\bxi=\sin \theta {\rm d}\theta {\rm d}\phi$.
The convolution of a bounded axisymmetric function $H(\bxi)=h(\cos \theta)$ with an arbitrary spherical function $X$ is well defined through
\begin{equation}
H \ast X(\bxi)=\int_\mathbb{S} h(\bxi\cdot \bxi^\prime)X(\xi^\prime){\rm d}\bxi^\prime\,.
\label{eq:apdix_convol_axisym}
\end{equation}
Meanwhile, the convolution theorem holds:
\begin{eqnarray}
H \ast X &=&\sum_{\ell m} h_{\ell} a_{\ell m} Y_{\ell m}
=\sum_{\ell m} h_{\ell} \langle Y_{\ell m}, X \rangle Y_{\ell m} \nonumber \\
&=& \sum_{j\in \mathcal{J}} \sum_{\ell}h_{\ell}^{(j)} \sum_{m=-\ell}^{\ell} \langle Y_{\ell m}, X \rangle Y_{\ell m} \nonumber \\
&=& \sum_{j\in \mathcal{J}}\sum_{\ell} h_{\ell}^{(j)} \Pi_{\ell} X \nonumber\\
&=& \sum_{j\in \mathcal{J}} \Psi^{(j)} X\,,
\label{eq:apdix_convol_axisym_multi}
\end{eqnarray}
where $a_{\ell m}$ are the multipole moments of $X$, $h_{\ell}$ are the Legendre series coefficients of $h$, $\mathcal{J}$ is a countable index set and $\Pi_{\ell}$ and $\Psi^{(j)}$ are operators defined as follows:
\begin{eqnarray}
\Pi_{\ell}X(\bxi)&=& \sum_{m=-\ell}^{\ell} \langle Y_{\ell m}, X\rangle Y_{\ell m}(\bxi)
=\frac{2\ell+1}{4 \pi}\int_\mathbb{S} X(\bxi^\prime)P_\ell(\bxi \cdot \bxi^\prime) {\rm d}\bxi^\prime\,,
\label{eq:apdix_Pi_ell}\\
\Psi^{(j)}&=&\sum_{\ell =0}^{\infty} h_{\ell}^{(j)} \Pi_{\ell}\,,
\label{eq:apdix_Psi_j}
\end{eqnarray}
where $P_{\ell}$ functions are the Legendre polynomials.
With Eqs.~\eqref{eq:apdix_convol_axisym} and~\eqref{eq:apdix_convol_axisym_multi}, isotropic wavelet analysis can be implemented either in the spatial domain or in the harmonic domain.
Here, we choose the latter by multiplying the harmonic coefficients of the field of interest $X$ with a spectral window $h_{\ell}$.
We call this the exact reconstruction condition that ensures $\sum_{j \in \mathcal{J}}\Psi^{(j)}=\textbf{Id}$, or equivalently
\begin{equation}
\sum_{j\in \mathcal{J}} h_{\ell}^{(j)}= 1\,.
\end{equation}

The filter function used in this paper is defined as follows~\citep{Pietrobon:2010jg,Marinucci:2007aj},
\begin{eqnarray}
\rho(t)&=&\left\{ \begin{array}{ll}
e^{-1/(1-t^{2})}\,,~ & -1< t < 1 \,,\\
0\,,~ & \text{otherwise} \,,
\end{array}
\right.
\\
s(u) &=& \left[\int_{-1}^u \rho(t) \,{\rm d} t \right]\Big/ \left[\int_{-1}^{1} \rho(t) {\rm d} t \right] \\
\varphi(x) &=& \left\{\begin{array}{ll}
1\,,~ & x\le 1/B\,, \\
s\left[1-2B\left(x-1/B \right)/(B-1) \right] \,,~ &1/B< x < 1 \,, \\
0\,,~ &x\ge 1
\end{array}
\right.
\\
\phi(x)&=& \varphi\left(\frac{x}{B} \right) - \varphi(x)\,.
\end{eqnarray}
Each function in the window family is $b_{\ell}^{(j)}=\sqrt{\phi(\ell/B^j)}\,, j\in \mathcal{J}$, complemented by $b_{\ell}^{(-1)}=\delta_{0 \ell}$. 
For the band-limited {\tt HEALPix} map with $\ell\leq\ell_\mathrm{max}$, the domain $\mathcal{J}$ is $[0, \log_B (\ell_\mathrm{max}) +1]$.
The window $b_{\ell}^{(j)}$ only overlaps with adjacent windows $b_{\ell}^{(j-1)}$ and $b_{\ell}^{(j+1)}$.
In this case, the exact reconstruction condition for $h_{\ell}^{(j)}=\left[b_{\ell}^{(j)}\right]^2$, $\sum_{j\in \mathcal{J}} h_{\ell}^{(j)}= 1$ is satisfied.
Figures~\ref{fig:apdix_phi_example_gnilc} and ~\ref{fig:apdix_point_B_adic} give an example of window family with $B=1.7$ in the harmonic domain and the localization features in the spatial domain, respectively.

\section{Calculation of the eGNILC Bias and AIC with $m$-modes of Foreground and $N_{\rm p}$ Pixel Data}
\label{app:AIC}
We need to calculate the covariance matrix of the reconstructed 21-cm signal ($\hat{\bs}$) before obtaining the AIC value of the likelihood. From Eq.~(\ref{eq:R_hat_s_dof}), we need to calculate $\boldsymbol{C}_{\boldsymbol{s\delta}}$ which is equal to ${\rm E}(\boldsymbol{\delta} \boldsymbol{s}^T)$. Therefore, we need to substitute Eq.~(\ref{eq:Delta_xp}) into Eqs.~(\ref{eq:d_p1})-(\ref{eq:d_p4}), and utilize them to calculate the expectation value. Because in Eq.~(\ref{eq:d_p1}), the $\boldsymbol{f}$ is in the first order, its ensemble average with $\boldsymbol{\delta}$ is zero, but the Eqs.~(\ref{eq:d_p2})-(\ref{eq:d_p4}) will not give zero expectation because $\bDelta_{\bx}$ also contains first order $\boldsymbol{f}$ and $\bs$ (Eq.~(\ref{eq:Delta_xp})). Therefore, we have
\begin{eqnarray}
\boldsymbol{C}_{\boldsymbol{s\delta}} &=& {\rm E}\left(\boldsymbol{\delta} \boldsymbol{s}^T \right) \nonumber \\
&=& \frac{1}{N_p}\sum_q \mathrm{E} \left[
\bS \left(\bS^\mathrm{T}\bR^{-1}_{\boldsymbol{x}}\bS\right)^{-1}
\left(\bS^\mathrm{T}\bR^{-1}_{\boldsymbol{x}} \left(\bs_q \boldsymbol{f}_q^\mathrm{T}  + \boldsymbol{f}_q \bs_q^\mathrm{T}\right) \bR^{-1}_{\boldsymbol{x}}\bS\right)
\left(\bS^\mathrm{T}\bR^{-1}_{\boldsymbol{x}}\bS\right)^{-1} \bS^\mathrm{T} \bR^{-1}_{\boldsymbol{x}}\! \boldsymbol{f}_p \bs^\mathrm{T}_p \right.
\nonumber \\
&& - \left. \bS \left(\bS^\mathrm{T}\bR^{-1}_{\boldsymbol{x}}\bS\right)^{-1} \bS^\mathrm{T} \bR^{-1}_{\boldsymbol{x}} \left(\bs_q \boldsymbol{f}_q^\mathrm{T} + \boldsymbol{f}_q \bs_q^\mathrm{T}\right) \bR^{-1}_{\boldsymbol{x}}\! \boldsymbol{f}_p \bs^\mathrm{T}_p \right] \nonumber \\
&=& \frac{1}{N_p} \left[
\bS \left(\bS^\mathrm{T}\bR^{-1}_{\boldsymbol{x}}\bS\right)^{-1} \bS^\mathrm{T}\bR^{-1}_{\boldsymbol{x}} \Tr\left(
 \bR^{-1}_{\boldsymbol{x}}\bS \left(\bS^\mathrm{T}\bR^{-1}_{\boldsymbol{x}}\bS\right)^{-1} \bS^\mathrm{T} \bR^{-1}_{\boldsymbol{x}} \bR_{\vf} \right) \bR_{\bs} \right.  \nonumber \\
&& + \left. \bS \left(\bS^\mathrm{T}\bR^{-1}_{\boldsymbol{x}}\bS\right)^{-1} \bS^\mathrm{T}\bR^{-1}_{\boldsymbol{x}} \boldsymbol{f}_p \boldsymbol{s}_p^\mathrm{T}  \bR^{-1}_{\boldsymbol{x}}\bS \left(\bS^\mathrm{T}\bR^{-1}_{\boldsymbol{x}}\bS\right)^{-1} \bS^\mathrm{T} \bR^{-1}_{\boldsymbol{x}} \boldsymbol{f}_p \bs^\mathrm{T}_p \right. \nonumber  \\
&& - \left. \bS \left(\bS^\mathrm{T}\bR^{-1}_{\boldsymbol{x}}\bS\right)^{-1} \bS^\mathrm{T}\bR^{-1}_{\boldsymbol{x}} \Tr\left( \bR^{-1}_{\boldsymbol{x}} \bR_{\vf} \right) \bR_{\bs}  - \bS \left(\bS^\mathrm{T}\bR^{-1}_{\boldsymbol{x}}\bS\right)^{-1} \bS^\mathrm{T}\bR^{-1}_{\boldsymbol{x}} \boldsymbol{f}_p \bs_p^\mathrm{T} \bR^{-1}_{\boldsymbol{x}} \boldsymbol{f}_p \bs^\mathrm{T}_p \right] \nonumber \\
& \equiv & \boldsymbol{C}^{(1)}_{\boldsymbol{s\delta}} + \boldsymbol{C}^{(2)}_{\boldsymbol{s\delta}} + \boldsymbol{C}^{(3)}_{\boldsymbol{s\delta}}, \label{eq:cov_err_p4}
\end{eqnarray}
where, in the third equality, we have used the property of the expectation value ${\rm E} \left(\bs^{T}_{p}\bs_{q} \right) = \bR_{\bs} \delta_{pq}$. We have also used the matrix property (see also~\citealt{Hanson2009})
\begin{eqnarray}
    \left\langle \boldsymbol{x}^{T}A\boldsymbol{x} \right\rangle &=& \Tr \left[A \left\langle \boldsymbol{x}\boldsymbol{x}^{T} \right\rangle  \right]=
    \Tr \left[A C  \right],
\end{eqnarray}
where $C$ is the covariance matrix of the Gaussian random variable vector $\boldsymbol{x}$, and $A$ is any matrix that is multiplicable to the data vector $\boldsymbol{x}$. The sum over $q$ will take only the $p=q$ term out of the summation.

The identities $\bs = \bS \boldsymbol{t}$ and $\bR_{\bs} = \bS \bR_{\vt} \bS^\mathrm{T}$ can help us simplify Eq.~(\ref{eq:cov_err_p4}) further.
We can substitute $\bR_{\vf} = \bR_{\bx} - \bR_{\bs}$ and obtain the first part of the covariance matrix ($\boldsymbol{C}^{(1)}_{\boldsymbol{s\delta}}$) as\footnote{Here we neglect the front $1/N_{\rm p}$ factor for now and will recover it later.}
\begin{eqnarray}
\boldsymbol{C}_{\boldsymbol{s\delta}}^{(1)} &=& 
\bS \left(\bS^\mathrm{T}\bR^{-1}_{\boldsymbol{x}}\bS\right)^{-1} \bS^\mathrm{T}\bR^{-1}_{\boldsymbol{x}} \Tr\left(
 \bR^{-1}_{\boldsymbol{x}}\bS \left(\bS^\mathrm{T}\bR^{-1}_{\boldsymbol{x}}\bS\right)^{-1} \bS^\mathrm{T} \bR^{-1}_{\boldsymbol{x}} \bR_{\vf} \right) \bR_{\bs} \nonumber \\
&=& 
\bS \left(\bS^\mathrm{T}\bR^{-1}_{\boldsymbol{x}}\bS\right)^{-1} \bS^\mathrm{T}\bR^{-1}_{\boldsymbol{x}} \Tr\left(
 \bR^{-1}_{\boldsymbol{x}}\bS \left(\bS^\mathrm{T}\bR^{-1}_{\boldsymbol{x}}\bS\right)^{-1} \bS^\mathrm{T} \bR^{-1}_{\boldsymbol{x}} \left(\bR_{\boldsymbol{x}}-\bR_{\bs} \right) \right) \bR_{\bs} \nonumber \\
&=& 
\bS \left(\bS^\mathrm{T}\bR^{-1}_{\boldsymbol{x}}\bS\right)^{-1} \bS^\mathrm{T}\bR^{-1}_{\boldsymbol{x}} \left[\Tr\left(
 \bR^{-1}_{\boldsymbol{x}}\bS \left(\bS^\mathrm{T}\bR^{-1}_{\boldsymbol{x}}\bS\right)^{-1} \bS^\mathrm{T} \right) -
 \Tr\left(
 \bR^{-1}_{\boldsymbol{x}}\bS \left(\bS^\mathrm{T}\bR^{-1}_{\boldsymbol{x}}\bS\right)^{-1} \bS^\mathrm{T} \bR^{-1}_{\boldsymbol{x}}\bR_{\bs}  \right) 
 \right]\bR_{\bs}\nonumber \\
&=& \bS \left(\bS^\mathrm{T}\bR^{-1}_{\boldsymbol{x}}\bS\right)^{-1} \bS^\mathrm{T}\bR^{-1}_{\boldsymbol{x}} \left[ \Tr(\boldsymbol{I}_{n_\mathrm{ch}-m})  - \Tr\left(
 \bR^{-1}_{\boldsymbol{x}}\bS \left(\bS^\mathrm{T}\bR^{-1}_{\boldsymbol{x}}\bS\right)^{-1} \bS^\mathrm{T} \bR^{-1}_{\boldsymbol{x}} \bS \bR_{\vt} \bS^\mathrm{T} \right) \right] \bR_{\bs} \nonumber \\
&=& \bS \left(\bS^\mathrm{T}\bR^{-1}_{\boldsymbol{x}}\bS\right)^{-1} \bS^\mathrm{T}\bR^{-1}_{\boldsymbol{x}} \left[ (n_\mathrm{ch} -m) - \Tr\left(\bR^{-1}_{\boldsymbol{x}} \bR_{\bs} \right) \right] \bR_{\bs},\label{eq:Csd_1}
\end{eqnarray}
where, in the fourth equality, we have used the property that $\bS$ and $\bR_{\boldsymbol{x}}$ matrices are of the dimensions $n_{\rm ch}\times (n_{\rm ch}-m)$ and $n_{\rm ch}\times n_{\rm ch}$ respectively (Eqs.~(\ref{eq:eigen}) and (\ref{eq:Smat})).

We now calculate the second term of the covariance matrix. By using $\bs = \bS \boldsymbol{t}$ and $\bR_{\bs} = \bS \bR_{\vt} \bS^\mathrm{T}$, we have
\begin{eqnarray}
\boldsymbol{C}_{\boldsymbol{s\delta}}^{(2)} &=&  
 \bS \left(\bS^\mathrm{T}\bR^{-1}_{\boldsymbol{x}}\bS\right)^{-1} \bS^\mathrm{T}\bR^{-1}_{\boldsymbol{x}} \boldsymbol{f}_p \boldsymbol{s}_p^\mathrm{T}  \bR^{-1}_{\boldsymbol{x}}\bS \left(\bS^\mathrm{T}\bR^{-1}_{\boldsymbol{x}}\bS\right)^{-1} \bS^\mathrm{T} \bR^{-1}_{\boldsymbol{x}} \boldsymbol{f}_p \bs^\mathrm{T}_p
\nonumber \\
&=&  
 \bS \left(\bS^\mathrm{T}\bR^{-1}_{\boldsymbol{x}}\bS\right)^{-1} \bS^\mathrm{T}\bR^{-1}_{\boldsymbol{x}} \boldsymbol{f}_p \boldsymbol{t}_p^\mathrm{T} \bS^{\rm T}  \bR^{-1}_{\boldsymbol{x}}\bS \left(\bS^\mathrm{T}\bR^{-1}_{\boldsymbol{x}}\bS\right)^{-1} \bS^\mathrm{T} \bR^{-1}_{\boldsymbol{x}} \boldsymbol{f}_p \bs^\mathrm{T}_p
\nonumber \\
&=&  
 \bS \left(\bS^\mathrm{T}\bR^{-1}_{\boldsymbol{x}}\bS\right)^{-1} \bS^\mathrm{T}\bR^{-1}_{\boldsymbol{x}} \boldsymbol{f}_p \boldsymbol{t}_p^\mathrm{T} \bS^\mathrm{T} \bR^{-1}_{\boldsymbol{x}} \boldsymbol{f}_p \bs^\mathrm{T}_p
\nonumber \\
&=& \bS \left(\bS^\mathrm{T}\bR^{-1}_{\boldsymbol{x}}\bS\right)^{-1} \bS^\mathrm{T}\bR^{-1}_{\boldsymbol{x}} \boldsymbol{f}_p \bs_p^\mathrm{T} \bR^{-1}_{\boldsymbol{x}} \boldsymbol{f}_p \bs^\mathrm{T}_p \,,
\end{eqnarray}
which cancels the second term of $\boldsymbol{C}_{\boldsymbol{s\delta}}^{(3)}$ in Eq.~\eqref{eq:cov_err_p4}. Substituting $\bR_{\vf} = \bR_{\bx} - \bR_{\boldsymbol{s}}$ into the first term of $\boldsymbol{C}_{\boldsymbol{s\delta}}^{(3)}$, it becomes
\begin{eqnarray}
\boldsymbol{C}_{\boldsymbol{s\delta}}^{(3)} 
&=& 
-  \bS \left(\bS^\mathrm{T}\bR^{-1}_{\boldsymbol{x}}\bS\right)^{-1} \bS^\mathrm{T}\bR^{-1}_{\boldsymbol{x}} \Tr\left[ \bR^{-1}_{\boldsymbol{x}} \left(\bR_{\bx} - \bR_{\boldsymbol{s}} \right) \right] \bR_{\bs}
\nonumber \\
&=& - \bS \left(\bS^\mathrm{T}\bR^{-1}_{\boldsymbol{x}}\bS\right)^{-1} \bS^\mathrm{T}\bR^{-1}_{\boldsymbol{x}} \times \left[n_\mathrm{ch} - \Tr\left(\bR^{-1}_{\boldsymbol{x}} \bR_{\bs} \right) \right] \bR_{\bs} . \label{eq:Csd_3}
\end{eqnarray}

Combining Eqs.~(\ref{eq:Csd_1}) and~(\ref{eq:Csd_3}), we obtain the total covariance as
\begin{equation}
\boldsymbol{C}_{\boldsymbol{s\delta}} = -\frac{m}{N_p}\bR_{\bs}.
\end{equation}
Therefore, from Eq.~(\ref{eq:R_hat_s_dof}) the covariance of $\hat{\bs}$ becomes
\begin{eqnarray}
\hat{\bR}_{\bs}= \left(1 -2 \frac{m}{N_p}\right) \bR_{\bs}\,.
\end{eqnarray}
This equation shows that the eGNILC bias is a function of the foreground modes number $m$ and the effective domain size $N_p$. Finally, we obtain the modified AIC by taking the logarithmic likelihood of both the data and the 21-cm prior, 
\begin{eqnarray}
\mathrm{AIC}(m, N_\mathrm{p})= 2 m + n_\mathrm{ch}\left[\frac{1}{1-2m/N_{\rm p}} + \ln\left(1-2\frac{m}{N_\mathrm{p}}\right)\right]  + \sum_{i=1}^{n_\text{ch}-m}\left[\mu_i-\ln{\mu_i}-1\right] \nonumber ,
\end{eqnarray}
which is Eq.~(\ref{eq:AIC_mod}).
\bibliography{gnilc_ref}

\end{document}